\documentclass{emulateapj}
\usepackage{apjfonts}

\newcommand{\kms}{\mbox{ km s$^{-1}$ }}
\newcommand{\kmsend}{\mbox{ km s$^{-1}$}}

\newcommand{\cmthree}{\mbox{cm$^{-3}$}}
\newcommand{\cmtwo}{\mbox{cm$^{-2}$}}

\newcommand{\msun}{\mbox{ M$_{\sun}$ }}
\newcommand{\msunend}{\mbox{ M$_{\sun}$}}

\newcommand{\msunyr}{\mbox{ M$_\sun$yr$^{-1}$ }}
\newcommand{\msunyrend}{\mbox{ M$_\sun$yr$^{-1}$}}

\newcommand{\lsun}{\mbox{L$_{\sun}$}}
\newcommand{\lir}{\mbox{L$_{\rm IR}$ }}
\newcommand{\htwo}{\mbox{H$_2$}}
\newcommand{\kkms}{\mbox{K-km s$^{-1}$}}
\newcommand{\z}{\mbox{$z$}}
\newcommand{\zsim}{\mbox{$z \sim $}}

\newcommand{\magorrian}{\mbox{$M_{\rm BH}$-$M_{\rm bulge}$ }}

\shorttitle{Effect of Galactic Winds on CO Emission in Galaxy Mergers}
\shortauthors{Narayanan et al.}
 
\begin{document}
\title{The Role of Galactic Winds on Molecular Gas Emission from
Galaxy Mergers}

\author{Desika Narayanan\altaffilmark{1,6}, Thomas
J. Cox\altaffilmark{2}, Brandon Kelly\altaffilmark{1}, Romeel
Dav\'{e}\altaffilmark{1}, Lars Hernquist\altaffilmark{2}, Tiziana Di
Matteo\altaffilmark{5}, Philip F. Hopkins\altaffilmark{2}, Craig
Kulesa\altaffilmark{1}, Brant Robertson\altaffilmark{3,4}, Christopher
K. Walker\altaffilmark{1}}

\altaffiltext{1}{Steward Observatory,
University of Arizona, 933 N Cherry Ave, Tucson, Az, 85721, USA}

\altaffiltext{2}{Harvard-Smithsonian Center for Astrophysics, 
60 Garden Street, Cambridge, MA 02138, USA}

\altaffiltext{3}{Kavli Institute for Cosmological Physics and
Department of Astronomy and Astrophysics, University of Chicago, 933
East 56th St., Chicago, Il,, 60637}

\altaffiltext{4}{Spitzer Fellow}

\altaffiltext{5}{Carnegie Mellon University, 
Department of Physics, 5000 Forbes Ave., Pittsburgh, PA 15213}

\altaffiltext{6} {dnarayanan@as.arizona.edu} \slugcomment{Submitted to
ApJ, September 18th, 2007, Accepteed Jan 21st, 2008}

\begin{abstract}
Galactic winds from starbursts and Active Galactic Nuclei (AGN) are
thought to play an important role in driving galaxies along the
starburst-AGN sequence. Here, we assess the impact of these winds on
the CO emission from galaxy mergers, and, in particular, search for
signatures of starburst and AGN-feedback driven winds in the simulated
CO morphologies and emission line profiles. We do so by combining a 3D
non-LTE molecular line radiative transfer code with smoothed particle
hydrodynamics (SPH) simulations of galaxy mergers that include
prescriptions for star formation, black hole growth, a multiphase
interstellar medium (ISM), and the winds associated with star
formation and black hole growth. Our main results are: (1) Galactic
winds can drive outflows of masses $\sim$10$^8$-10$^9$\msun which may
be imaged via CO emission line mapping. (2) AGN feedback-driven winds
are able to drive detectable CO outflows for longer periods of time
than starburst-driven winds owing to the greater amount of energy
imparted to the ISM by AGN feedback compared to star formation. (3)
Galactic winds can control the spatial extent of the CO emission in
post-merger galaxies, and may serve as a physical motivation for the
sub-kiloparsec scale CO emission radii observed in local advanced
mergers. (4) Secondary emission peaks at velocities greater than the
circular velocity are seen in the CO emission lines in all models,
regardless of the associated wind model. In models with winds,
however, these high velocity peaks are seen to preferentially
correspond to outflowing gas entrained in winds, which is not the case
in the model without winds. The high velocity peaks seen in models
without winds are typically confined to velocity offsets (from the
systemic) $\la$1.7 times the circular velocity, whereas the models
with AGN feedback-driven winds can drive high velocity peaks to
$\sim$2.5 times the circular velocity.

\end{abstract}

\keywords{galaxies: active,  starburst, ISM -- quasars: general --
line: profiles -- ISM: jets and outflows}

\section{Introduction}

Observed relationships in galaxies between central black hole mass and
stellar mass (e.g. Magorrian et al. 1998), velocity dispersion
(i.e. the $M_BH$-$\sigma$ relation; e.g. Gebhardt et al. 2000;
Ferrarese \& Merritt 2000), or galaxy structural properties (e.g. the
black hole fundamental plane: Hopkins et al. 2007a,b) indicate a
co-eval nature in supermassive black hole growth and star formation in
galaxies.  These results have prompted a number of investigations in
recent years to quantify this apparent self-regulation in star
formation and black hole growth in galaxies, and their relationship to
the formation and evolutionary history of the host system
(e.g. Kauffmann \& Haehnelt, 2000; Hopkins et al. 2006a,b; 2007c,d).

Over the last two decades, observations of local galaxies have painted
a compelling picture in which galaxy mergers provide the link between
massive starbursts and central black hole growth and activity.  ULIRGs
(Ultraluminous Infrared Galaxies), for example, are a class of
starburst galaxies with elevated infrared luminosities (\lir $\geq$
10$^{12}$\lsun) which typically show signs of interactions
(e.g. Downes \& Solomon, 1998; Sanders et al. 1988a; Scoville et
al. 2000). While the intense infrared luminosity in these sources
certainly owes in large part to the merger induced starbursts
(e.g. Sanders et al. 1988a,b; Sanders \& Mirabel, 1996), in some cases
a contribution from a buried active galactic nucleus (AGN) may be
non-negligible.  For example, many ULIRGs show spectral energy
distributions (SEDs), infrared color ratios (e.g. F[25/60$\mu$m]),
polycyclic aromatic hydrocarbon emission (PAH) deficits, and emission
line fluxes (e.g. [Ne V] at 14.3$\mu$m) consistent with central AGN
activity (e.g. Armus et al. 2004, 2006; Farrah et al. 2003; de Grijp
et al. 1985).  Indeed, 35-50\% of ULIRGs with luminosity above \lir of
10$^{12.3} L_{\sun}$ show optical and NIR spectra consistent with AGN
activity (Kim, Veilleux \& Sanders, 2002; Tran et al. 2001; Veilleux,
Kim \& Sanders, 1998). These results suggest that these merging
systems are simultaneously undergoing a massive star formation and
central black hole growth phase.

Observed similarities such as these between starburst galaxies, ULIRGs
and quasars prompted Soifer et al. (1987) and Sanders et al.
(1988a,b) to propose an empirically derived evolutionary sequence
which connects galactic starbursts to quasars through galaxy
mergers. In this picture, the fueling of the central black hole and
AGN growth phase is realized during the major merger when gaseous
inflows trigger nuclear starbursts as well as central black hole
growth. The dusty galaxy transitions from a cold, starburst dominated
ULIRG, to a warm, AGN-dominated ULIRG (where 'cold' and 'warm' refer
to F[25/60$\mu$m] ratios), and, as supernovae and stellar winds clear
the obscuring gas and dust, to an optical quasar. In this sense,
galaxy mergers provide a unique laboratory for studying the possible
co-evolution of starbursts, black hole growth and activity, and
spheroid formation.

Recent models have provided a theoretical foundation and further
evidence for a merger-driven starburst-AGN connection in galaxies.
Specifically, simulations by Springel, Di Matteo \& Hernquist (2005a)
have shown that galaxy mergers can fuel large-scale gaseous inflows
(e.g. Barnes \& Hernquist, 1991, 1996) which trigger nuclear
starbursts (Mihos \& Hernquist, 1996; Springel et al. 2005a) as well
as promote central black hole growth (Di Matteo, Springel \&
Hernquist, 2005). Subsequent winds associated with the growth of
central black holes can lift the veil of obscuring gas and dust, and
along several sightlines produce a quasar with comparable lifetimes,
luminosity functions, and observed $B$-band and X-ray properties to
those observed (Cox et al., 2006b; Hopkins et al. 2005a-d; 2006a-d).
The merger remnants quickly redden owing to gas depletion and the
impact of feedback from star formation and black hole growth (e.g.
Springel et al. 2005b) and resemble elliptical galaxies in their
kinematic and structural properties.  Indeed, the population of stars
formed during the starbursts accounts for the central ``excess light''
seen in ongoing mergers (e.g. Rothberg \& Joseph 2004, 2006) and
provides a detailed explanation for the luminosity profiles of old
ellipticals (e.g.  Kormendy et al. 2007; see Mihos \& Hernquist 1994a;
Hopkins et al. 2007e,f,g).

A consensus picture has thus been borne out from these observations
and simulations over the last two decades in which so called
'feedback' processes associated with winds from star formation and
black holes act to self-regulate the growth of both the stellar and
black hole masses in galaxies (e.g. Fabian, 1999; Silk \& Rees,
1998). Indeed, the effects of starburst and AGN feedback-driven winds
have been observed in local galaxies (e.g. Heckman et al. 2000;
Martin, 2005; Rupke, Veilleux \& Sanders, 2005a-c; Rupke \& Veilleux,
2005; Tremonti, Moustakas \& Diamond-Stanic, 2007), as well as those
at high-\z \ (e.g. Narayanan et al. 2004; Pettini et al. 2002; Shapley
et al. 2003) by way of absorption line outflows. However, the direct
effect of feedback processes (especially from the highly efficient
central AGN) on the emission properties of galaxies is not yet well
characterized (see Veilleux, Cecil \& Bland-Hawthorn, 2005 for an
extensive review and associated references). In this sense, it is
important to relate theories of galaxy formation and evolution which
incorporate physically motivated models of feedback processes to
observational signatures of winds across the electromagnetic spectrum.

In particular, observations of molecular gas in galaxies have proven
valuable in characterizing the physics related to nuclear star
formation and central AGN fueling in galaxy mergers as the molecular
gas serves as fuel both for star formation as well as the central
black hole(s).  For example, interferometric observations of molecular
gas in ULIRGs show that most typically harbor $\sim$10$^{10}$\msun of
molecular gas within the central 1.5 kpc (Scoville et al. 1986; Bryant
\& Scoville, 1999). Moreover, high resolution maps of dense molecular
gas at submillimeter wavelengths in ULIRGs have revealed kinematic
structures of double nuclei in mergers (Scoville, Yun \& Bryant,
1997), bar-driven inflows (Sakamoto et al. 2004), and the density
structure of gas fueling the central AGN (Iono et al. 2004).

With the increased spatial resolution and sensitivity afforded by the
latest generation of (sub)mm-wave interferometers (e.g. the SMA,
CARMA, PdBI), a number of recent observations have been able to
pioneer investigations as to the effects of galactic winds on
molecular gas emission in galaxies (e.g. Iono et al. 2007; Sakamoto,
Ho \& Peck, 2006; Walter, Wei\ss \ \& Scoville, 2002). Detections such
as these are expected to become more routine in upcoming years as the
ALMA interferometer becomes available. An important complement to
these current and forthcoming observations of molecular line emission
from starburst galaxies and AGN are physical models which directly
relate CO emission properties to galactic scale winds.

In this context, it is our aim to investigate the role that galactic
winds can play on CO emission properties from starburst galaxies and
AGN via numerical simulations. In particular, we focus on specific
signatures of winds imprinted on CO morphologies and emission line
profiles. In this work, we present self-consistent radiative transfer
calculations for the emission properties of CO molecular gas in
gas-rich galaxy mergers which account for the winds associated with
both star formation and black hole growth.

In \S~\ref{section:numericalmethods}, we describe the hydrodynamic
simulations, and radiative transfer methodology. In
\S~\ref{section:spiral} we provide an example of our methods by
applying our radiative transfer calculations to a model of a
star-forming disk galaxy.  In \S~\ref{section:morphology}, we discuss
the effect of winds on observed CO morphologies. We explore the
response of emission line profiles to winds in
\S~\ref{section:lineprofiles}, present a broader discussion of these
results with respect to observations in \S~\ref{section:observations},
and summarize in \S~\ref{section:conclusions}. Throughout this paper,
we assume a $\Lambda$CDM cosmology with $h$=0.7, $\Omega_\Lambda$=0.7,
$\Omega_{\rm M}$=0.3.

\section{Methodology}

Simulating molecular line emission on galaxy-wide scales has a
relatively short history owing to the computational costs associated
with the hydrodynamics and radiative transfer. Early works by Silk \&
Spaans (1997) and Combes, Maoli \& Omont (1999) focused on simulating
molecular emission from high-\z \ objects, using idealized galaxy
systems. Wada \& Tomisaka (2005) pioneered incorporating
three-dimensional non local thermodynamic equilibrium (non-LTE)
radiative transfer calculations into hydrodynamic simulations of AGN,
focusing specifically on predicted emission from the circumnuclear
molecular torus. More recent works have folded non-LTE radiative
transfer codes into self-consistent hydrodynamic simulations of galaxy
mergers. These works have been performed in mergers scaled for low
redshift (Narayanan et al. 2006a), as well as those appropriate for
higher-\z \ systems (Greve \& Sommer-Larsen, 2006; Narayanan et
al. 2007). Here, we employ a similar methodology to that of Narayanan
et al. (2006a, 2007). In this section, we describe our methods for
incorporating 3D non-LTE molecular line radiative transfer
calculations into hydrodynamic simulations of galaxy mergers which
account for a multi-phase ISM, star formation, growing black holes,
and the effects of winds associated with star formation and central
AGN.

\subsection{Numerical Methods}
\label{section:numericalmethods}

\subsubsection{Hydrodynamics}
The hydrodynamic simulations used for this study employed a modified
version of the publicly available $N$-body/SPH code, GADGET-2
(Springel, 2005), adopting the fully conservative formulation of SPH
developed by Springel \& Hernquist (2002).  The methods used to
construct the progenitor galaxies, as well as a detailed description
of the algorithms used to simulate the physics of feedback from star
formation and accreting black holes is detailed in Springel, Di Matteo
\& Hernquist (2005a), and we direct the reader toward this work for
further information. Here, we briefly summarize these methods, and
describe the specifics of the modeling most pertinent to this work.

 The interstellar medium (ISM) is modeled as a multi-phase medium in
which pressure feedback from supernovae heating is treated through an
effective equation of state (EOS) (Springel, Di Matteo \& Hernquist,
2005a). The disk galaxies used as progenitors in these simulations
employed a softened EOS with softening parameter $q_{\rm EOS}$=0.25
(see Figure 4 of Springel et al. 2005a). The ISM is modeled to include
cold, dense gas surrounded by a hot ISM, and is realized numerically
through ``hybrid'' SPH particles (Springel \& Hernquist, 2003). In
this formulation, cold ISM is allowed to grow through radiative
cooling of the hot ISM, and conversely feedback associated with star
formation can evaporate cold clouds into diffuse, hot gas. Star
formation follows the prescription of Springel \& Hernquist (2003) and
is constrained to fit the Schmidt/Kennicutt observed star formation
laws (Kennicutt, 1998a,b; Schmidt, 1959).

In order to explore the effects of galactic winds on the molecular
ISM, we include a formulation for both accreting black holes, and
winds associated with massive starbursts. In our SPH formalism, the
black holes are included as sink particles which accrete gas from the
surrounding ISM. The accretion is treated with a Bondi-Lyttleton-Hoyle
parametrization (Bondi \& Hoyle, 1944, Hoyle \& Lyttleton, 1939) with
a fixed maximum rate corresponding to the Eddington limit. The black
hole radiates such that its bolometric luminosity is given by
$L$=$\epsilon \dot{M} c^2$ with accretion efficiency
$\epsilon$=0.1. We further assume that 5\% of this energy couples to
the surrounding ISM such that 0.5\% of the accreted mass energy in our
simulations is reinjected isotropically into the ISM as thermal energy
(a parameter choice which allows the merged galaxy to reproduce the
local $M_BH$-$\sigma$ normalization [Di Matteo, Springel \& Hernquist,
2005]). Hence, the black hole energy deposition rate into the ISM may
be expressed as

\begin{equation}
\label{eq:bhenergy}
\dot{E}_{\rm BH} = \epsilon_{\rm effective} \dot{M}_{\rm BH} c^2
\end{equation}
where $\epsilon_{\rm effective}$=0.005 (Cox et al. 2007; Springel,
Di Matteo \& Hernquist., 2005a).

Starburst winds are treated utilizing the constant wind models of
Springel \& Hernquist (2003). In this formulation, the mass loss rate
(denoted by $\dot{M_w}$) is assumed to be proportional to the star
formation rate (denoted by $\dot{M_\star}$), such that:
\begin{equation}
\dot{M_w} = \eta\dot{M_\star}
\end{equation}
where $\eta$ quantifies the wind mass loading factor. The wind is
assumed to have a speed constrained by a fraction of the supernova
energy ($\chi$):
\begin{equation}
\frac{1}{2}\dot{M_w}v_w^2 = \chi \epsilon_{\rm SN}\dot{M_\star}
\end{equation}
In a recent study of galactic winds in galaxy mergers, Cox et al.
(2007) found that winds with high mass loading factors ($\eta \ge$1)
may be unphysical in that they tend to prevent any starburst after the
major merger. We thus choose a moderate mass loading factor of
$\eta$=0.5, and constant wind velocity $v_w$=837 \kms for this
study. These numbers are consistent with the favored momentum-driven
wind scalings for the galaxy masses we simulate found by Oppenheimer
\& Dav\'e (2006). For more details concerning constraints on wind
velocities and efficiencies in galactic scale simulations, we refer
the reader to Cox et al.  (2007).

The energy input rate into the ISM from star formation may be expressed as 
\begin{equation}
\label{eq:sbenergy}
\dot{E_{\rm sb}}= \epsilon_{\rm SN}\dot{M_\star}
\end{equation}
where $\epsilon_{\rm SN}$ is the energy imparted by supernovae, per
solar mass, taken here to be 1.4 $\times$10$^{49}$ ergs \msun$^{-1}$
(Springel \& Hernquist 2003).

 The progenitor galaxies in our models contain a dark matter halo
initialized with a Hernquist (1990) profile, concentration index
$c$=9, spin parameter $\lambda$=0.033, circular velocity $V_{200}$=160
\kmsend, and are bulge-less.  The galaxy's exponential disk is
rotationally supported and comprises 4.1\% of the total mass.  We
utilize 120,000 dark matter particles, and 80,000 total disk
particles, 40\% of which are gaseous in nature, the remainder serving
as collisionless star particles. The total masses of the progenitor
galaxies were 1.4$\times$10$^{12}$\msunend, and the final merger
produced a central black hole with mass
$\sim$5$\times$10$^7$\msunend. The softening lengths were 100 pc for
baryons, and 200 pc for dark matter.

We have conducted five binary merger simulations.  Each of the
progenitor galaxies in our simulations was constructed with the
aforementioned physical parameters, and the varied parameters include
initial orbit and types of wind included (e.g. starburst and/or
AGN). Identical progenitor galaxies were used in each simulation.  The
initial conditions for the merger simulations are summarized in
Table~\ref{table:ICs}, and we will henceforth refer to the simulations
by their model name listed in Column 1 of Table~\ref{table:ICs}. In
order to simplify analysis throughout this paper we will largely
analyze the properties of the four simulations which share the same
merger orientation and utilize the coplanar simulations (model co-BH)
for only specific comparisons.

\begin{deluxetable*}{cccccccccc}
\tabletypesize{\scriptsize}
\tablecaption{Progenitor Galaxies\label{table:ICs}}
\tablewidth{0pt}
\tablehead{
\colhead{Run} & \colhead{Other Names Used} & \colhead{$\theta_1$}  &
\colhead{$\phi_1$} & \colhead{$\theta_2$} &
\colhead{$\phi_2$} & \colhead{V$_{\rm vir}$} &
\colhead{Gas Frac} &
\colhead{BH} & \colhead{SNe}\\
&&&&&&(\kmsend)&&&
}
\startdata
BH & e     & 30  & 60 & -30 & 45 & 160 & 0.4 &  yes & no\\
no-winds& e-no & 30  & 60 & -30 & 45 & 160 & 0.4 &  no  & no\\
sb & --    & 30  & 60 & -30 & 45 & 160 & 0.4 &  no  & yes\\
sbBH& -- &30   & 60 & -30 & 45 & 160 & 0.4 &  yes & yes\\
co-BH & h & 0 & 0 & 0 & 0 & 160 & 0.4 & yes & no\\ \enddata
\tablenotetext{1}{Column 1 is the name of the model used in this work.
Column 2 lists alternative names for these models used by Chakrabarti
et al. (2007a,b), Cox et al. (2006a,b; 2007), Hopkins et al. (2005a-d;
2006a-d; 2007h), and Robertson et al. (2006a-c), for comparison. Columns 3 \&
4 are initial orientations for disk 1, Columns 5 \& 6 are for disk
2. Column 7 gives the virial velocity of the progenitors and Column 8
gives their initial gas fractions. Columns 9 and 10 describe the types
of winds (optionally) included in each model.}
\end{deluxetable*}

\subsubsection{Radiative Transfer}

Submillimeter and millimeter wave radiation from molecules is
dependent on the distribution of level populations which depend both
on collisions with other molecules and atoms, as well as the incident
radiation field.  The excitation of molecular gas is highly sensitive
to both variations in the temperature and density distribution, as
well as the incident radiation field. There are usually large
differences in the collisional densities necessary to excite different
energy levels. For example, in the case of CO, the critical density to
collisionally populate the J=1 state is $n_{\rm crit}
\sim$10$^2$-10$^3$ cm$^{-3}$ while the J=3 state typically requires
$n_{\rm crit} \sim$10$^4$ cm$^{-3}$.  The latter density is
characteristic of dense cores in GMCs, whereas the former is typical
of diffuse GMC atmospheres.

The non-LTE radiative transfer calculations for this work were
calculated during post-processing of the hydrodynamic simulations. The
spatial range considered for the radiative transfer was 12 kpc.  The
simulation outputs were smoothed to a resolution of $\sim$250 pc in
order to defray computational costs associated with the radiative
transfer simulations. In order to model the strongly density-dependent
excitation rates in our simulations, we have expanded the Bernes
(1979) non-local thermodynamic equilibrium (LTE) Monte Carlo radiative
transfer algorithm to include a mass spectrum of GMCs in a subgrid
manner.

The molecular gas fraction in galaxies is determined by the
metallicity, dust content, interstellar radiation field, density and
temperature (Hollenbach, Werner \& Salpeter, 1971; Pelupessy,
Papadopoulos \& van der Werf, 2006). However, owing to limited spatial
resolution, this calculation cannot be done explicitly as the location
of individual stars and clouds are not know.  We therefore assume that
half of the cold neutral gas mass in each grid cell is in atomic form
and half in molecular, as motivated by local volume surveys of star
forming galaxies (e.g. Keres, Yun \& Young 2003) though note that an
estimate of the molecular fraction can be obtained via sub-grid
prescriptions (e.g. Pelupessy et al. 2006; Greve \& Sommer-Larsen
2006; Robertson \& Kravtsov 2007).

In our formulation, the cold molecular gas is assumed to be exist in a
mass spectrum of GMCs in each grid cell
\begin{equation}
\frac{dN}{dM} \propto M^{-\beta}
\end{equation}
where we take $\beta$=1.8 (Blitz et al. 2006). The GMCs are modeled as
power-law spheres where the density ($n$) is given
by:
\begin{equation}
n=n_0\left[\frac{r_0}{r}\right]^\alpha
\end{equation}
The radius of the cloud is determined by the Galactic GMC mass-radius
relation (Rosolowsky 2005, 2007; Solomon et al. 1987), such that the
sum of the cloud density distributions in a grid cell takes the form:
\begin{equation}
n=\frac{1}{r^\alpha_m}\left[\sum_m r_{0,m}^\alpha n_{0,m}\right]
\label{eq:globaldist}
\end{equation}

which is used to set the initial level populations, as well as the
densities involved in the collisional rate calculations. For this work
we utilize a cloud power-law index of $\alpha$=2 (Walker, Adams \&
Lada 1990). Observational evidence suggests that a range of power law
indices in GMCs from $\alpha$=1-2 may be appropriate (Andre,
Ward-Thompson \& Motte 1996; Fuller \& Myers 1992; Ward-Thompson et
al. 1994), although our results are not very sensitive to the choice
of power law index within this range.

Formally, we build an emergent spectrum by integrating the equation of
radiative transfer along the line of sight:
\begin{equation}
  I_\nu = \sum_{z_0}^{z}S_\nu (z) \left [ 1-e^{-\tau_\nu(z)} \right
  ]e^{-\tau_\nu(\rm tot)}
\end{equation}
where $I_\nu$ is the frequency dependent intensity, $S_\nu$ is the
source function, $\tau$ is the optical depth, and $z$ is the position
along the line of sight.

If the level populations are in LTE, then the source function, $S_\nu$
can be simply replaced by the Planck function.  However, when
considering the propagation of lines through a medium with
insufficient density for collisions to thermalize the level
populations (e.g., $n\ll n_{\rm crit}$), the effects of radiative
excitation and de-excitation must be considered. In this case, the
source functions must be calculated explicitly.

The source function from each cloud, $m$, for a given transition from
upper level to lower level $u\rightarrow l$ is given by
  \begin{equation}
S_{\nu,m}=\frac{n_{u,m}A_{ul}}{(n_{l,m}B_{lu}-n_{u,m}B_{ul})}
\end{equation}
where the level populations are assumed to be in statistical equilibrium, and 
determined through the rate equations:
\begin{eqnarray}
 \nonumber n_{l,m}\left [\sum_{k<l}A_{lk}+\sum_{k\neq l}(B_{lk}J_{\nu}+
    C_{lk})\right]= \\ 
  \sum_{k>l}n_{k,m}A_{kl}+\sum_{k\neq l}n_{k,m}(B_{kl}J_\nu+C_{kl}) \, .
\end{eqnarray}
$A$, $B_{kl}$ and $B_{lk}$ are the Einstein rates for spontaneous
emission, stimulated emission, and absorption, respectively, $C$ are
the collisional rate coefficients, the indices $l$ and $k$ represent
different energy levels, and $J_\nu$ is the mean intensity through a
given grid cell:
\begin{equation}
J_\nu=\frac{1}{4\pi}\int I_\nu d\Omega
\end{equation}
As is evident by the previous three equations, the problem is
circular: the observed intensity depends on the source function which
is determined by the level populations. When in a non-LTE regime ($n
\ll n_{\rm crit}$), collisions alone do not determine the level
populations, but rather the mean intensity field (e.g. radiation from
gas in other grid cells) plays a role as well. The solution is
achieved by means of iteration.

The initial level populations in a grid cell are estimated based on
the global density distribution for all clouds in a grid cell
(Equation~\ref{eq:globaldist}). We emit model photons which represent
many real photons from the mass spectrum of GMCs in each grid
cell. The model photons are given a weight $W$ proportional to the
total number of molecules in the upper state of a given transition in
a cell, and the Einstein $A$ rate coefficient for the transition.  The
photons are emitted isotropically in 3 dimensions and have line
frequency randomly drawn from the line profile function:
\begin{equation}
\phi(\nu)=\frac{1}{\sigma \sqrt{\pi}}{\rm exp}\left \{-\left
  (\nu-\nu_0-\bf{v \cdot \hat n} \frac{\nu_{ul}}{c}\right )^2 /
  \sigma^2\right \}
\end{equation}
where the effects of the local kinetic temperature and microturbulent
velocity field are accounted for via 
\begin{equation}
\sigma=\frac{\nu_0}{c}\left [\frac{2kT}{m}+V_{\rm turb}\right ]^{\frac{1}{2}}
\end{equation}
The photon then takes a step over distance $s$ through the next grid
cell. After taking a step, the weight is attenuated by a factor
$e^{-\tau}$ where the opacities are given by
\begin{eqnarray}
\alpha_\nu(\rm dust)=\kappa_\nu \rho_{\rm dust}\\ 
\alpha_\nu^{ul}(\rm gas)= \frac{h \nu_{ul}}{4\pi}\phi(\nu)(n_lB_{lu}-n_uB_{ul})
\end{eqnarray}
and
\begin{equation}
\tau_{\nu,m}=(\alpha_{\nu,m}(\rm dust)+\alpha_{\nu,m}(\rm gas))\times s
\end{equation}

\begin{figure}
\includegraphics[angle=90,scale=0.35]{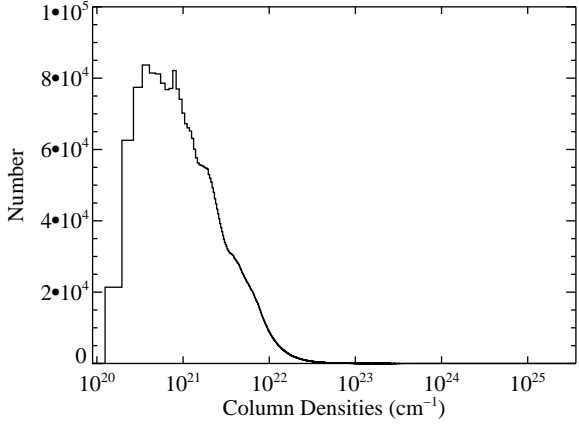}
\caption{Representative distribution of column densities for a grid
cell of mass $\sim$3$\times$10$^6$\msun. Distribution of columns is
derived by randomly placing clouds drawn from a mass spectrum until
the distribution converges.  Photons entering a grid cell see a column
randomly drawn from a distribution such as this in order to simulate
the actual column seen by a photon in a statistical manner.
\label{figure:column_dist}}
\end{figure}

The column seen by a photon in a grid cell is calculated via a Monte
Carlo approach. Specifically, we calculate the distribution of
potential column densities a photon could see through a given grid
cell by simulating a sample of the subgrid cells individually on
higher resolution (512$^3$) grids. In these higher resolution grids,
we place GMCs at random locations with masses drawn from the power-law
mass spectrum, and radii given by the Galactic mass-radius
relation. We do this numerous times until the distribution of
potential columns a photon entering at random angle sees
converges. The distribution of columns is found to scale in a
self-similar manner with grid cell (molecular) mass. We then
explicitly draw from this distribution to determine the column seen by
a model photon as it leaves and enters given grid cells. To illustrate
this, we plot an example of the distribution of column densities in
Figure~\ref{figure:column_dist}.  The photon continues to propagate
through grid cells in this manner until the weight is negligible, or
the photon has left the grid.

Once a sufficiently large number of model photons have been emitted to
simulate the mean radiation field, $J_\nu$, the level populations can
be updated via the rate equations. In the rate equations (also
commonly called the equations of statistical equilibrium), the terms
involving the mean intensity represent the radiatively induced
excitations (absorption) and stimulated emission in a given cell of
clouds $m'$. We may rewrite these equations by defining the number of
excitations of a molecule in lower state $l$ by
\begin{equation}
S_{lu,m'}=\frac{h \nu_{ul}}{4\pi}\phi(\nu)B_{lu}\frac{sW_0}{V_{m'} \tau}
(1-e^{-\tau(\rm tot)})
\end{equation}
where $V$ is the volume of the cell and $W_0$ is the original weight
of the photon.

Following Bernes (1979), we then rewrite the equations of statistical
equilibrium as
\begin{eqnarray}
 \nonumber n_l\left [\sum_{k<l}A_{lk}+\sum_{k\neq l}\left
  (S_{lk,m'}+C_{lk}\right )\right ]= \\
  \sum_{k>l}n_kA_{kl}+\sum_{k\neq
    l}n_k\left [\frac{g_l}{g_k}S_{lk,m'}+C_{kl}\right ]
  \label{eq:stateq2}
\end{eqnarray}
where $g$ are the statistical weights of the level. 

After a single generation of photons has been emitted, the $S_{lu,m'}$
 is calculated for each grid cell, $m'$. We can then use the analog of
 the previous equation to calculate the updated level populations for
 the $N$ individual radial cells that make up the subgrid spectrum of
 GMCs. The relative contribution of each of the $N$ cells to the total
 number of excitations ($S_{lu,m'}$) in a given grid cell is
 determined via the column across that region.

New level populations are calculated for each of the $N$ cells in the
spectrum of GMCs via matrix inversion and then summed to give the
total populations for their parent grid cell. New weights $W$ are
given to a new generation of emergent photons, and the process is
iterated upon until the level populations across the grid are
converged.  We typically emitted 12$\times$10$^6$ model photons per
iteration and the boundary conditions for the radiative transfer
included the 2.73K microwave background.

The radiative transfer and excitation depend sensitively on the
accuracy of the rate coefficients. We have obtained our coefficients
from the {\it Leiden Atomic and Molecular Database} (Schoier et
al. 2005). We have tested our radiative transfer codes per publicly
available benchmarks published in van Zadelhoff et al. (2002).  These
tests comprise ``inside-out'' collapsing spherical molecular clouds
and the details of the performance of our codes under these standard
tests can be found in Figure 1 of Narayanan et al. (2006b).

Owing to its relatively high abundance (CO/\htwo=1.5$\times$10$^{-4}$
in the Galaxy; Lee, Bettens \& Herbst, 1996) compared to other common
molecules (e.g. HCN, CS and HCO$^+$), $^{12}$CO (hereafter, CO) is the
most commonly observed molecular tracer of \htwo. We therefore focus
our modeling efforts for this work on the predicted CO emission
properties of galaxy mergers in the context of galactic winds, and
defer investigations of emission from other molecules to future work.

\subsubsection{Model Assumptions}

This methodology has the distinct advantages of being able to simulate
the effects of dense cores as well as diffuse atmospheres of
clouds. Because we divide the GMCs into a series of sub-grid cells, we
are not constrained to using a single density in our rate equations,
and thus the collisional excitation characteristic of dense cores can
be well-represented while the influence of radiation on more diffuse
regions can likewise be accounted for. Moreover, we are able to
include these features in our calculations without expanding the grid
to computationally prohibitive sizes.

Our subgrid formulation is dependent on a series of assumptions as
well. First, owing to limited spatial resolution, we are forced to
assume a constant \htwo \ gas fraction constrained by observational
surveys. This molecular gas in a given grid cell is assumed to be all
in GMCs.  The GMCs are realized as power-law spheres in our model
whereas clouds are understood observationally to be fractal in nature
(e.g. Elmegreen \& Falgarone, 1996).  The clouds within each grid cell
are all at the same temperature, and individually isothermal (though
temperatures are allowed to vary from grid cell to grid cell, and thus
temperature gradients exist across the galaxy).

The temperatures in the cold phase of the ISM in the hydrodynamic
models are fixed at 1000K (Springel \& Hernquist, 2003).  This choice
is arbitrary and has no effect on the hydrodynamic simulations.
Typically, temperatures in molecular clouds range from 10-100K;
temperatures of 1000K would systematically put too much weight on the
collisional coefficients in the rate equations, resulting in overly
excited level populations. While it is indeed tractable to run
galaxy-wide hydrodynamic models which allow the temperatures to cool
to lower values than those set here (e.g. Robertson \& Kravtsov 2007),
it is computationally infeasible for the large numbers of simulations
presented in this work (Table~\ref{table:ICs}). Thus, a subgrid model
for refining the temperature structure of the ISM is required.  We
construct a relatively simple model for the ISM temperature structure
by considering the dominant heating mechanism for the gas as heating
from O and B stars (assuming a Salpeter IMF). We sample individual
grid cells on a finer (512$^3$) resolution and place O and B stars
formed in the hydrodynamic simulations in the power-law sphere clouds
randomly throughout the higher resolution grid cells. These higher
resolution cells are utilized to calculate the mean column of gas and
dust (which are assumed to have the same spatial distribution in the
clouds) seen by the stellar UV flux, and the mean blackbody heating
rate is derived for each grid cell. When no O stars are present, lower
mass stars dominate the heating of the gas. Owing to the large grid
cell size ($\sim$250 pc), and the high densities of the molecular
cores in the subgrid clouds ($\gtrsim$10$^5$\cmthree), blister regions
from the few formed O stars have a negligible impact on the total
molecular gas content (Hollenbach \& Tielens 1999). The neutral gas is
further allowed to cool via metal line cooling which is calculated
using a mean escape probability radiative transfer code (ESCAPE;
Kulesa 2002; Kulesa et al. 2005).  The dominant ISM coolants
considered are CII, NII and atomic Oxygen and Carbon with abundances
set at Galactic values (Lee, Bettens \& Herbst 1996). Heating and
excitation from a central AGN (when included) is not considered, and
is deferred to a future study.

Utilizing this methodology, we find typical isothermal cloud
temperatures ranging from 10K-120K (nominally 10-30 K for simulations
scaled for the local Universe such as those presented in this work,
though temperatures can approach $\sim$120 K for models of higher
redshift galaxies; Narayanan et al. 2007a). It is additionally worth
noting as well that variations in the temperature structure of the ISM
within this range are ancillary with respect to determining the
excitation patterns of the molecular ISM. As an example, doubling the
temperature from 10K to 20K only changes the collisional rates of
excitation of e.g. CO J=2 to CO J=3 by $\sim$10\% (utilizing values
taken from the Leiden Atomic and Molecular Database; Schoier et
al. 2005). The dominant effect is instead the density structure of the
GMCs. Tests have shown that the excitation conditions of these models
are robust while the assumed cloud power-law index and/or GMC mass
spectrum index remain within observational constraints (Blitz et
al. 2006; Rosolowsky 2007; see also Narayanan et al. 2007b).

\subsubsection{Resolution Tests}

\begin{figure*}
\includegraphics[scale=0.9]{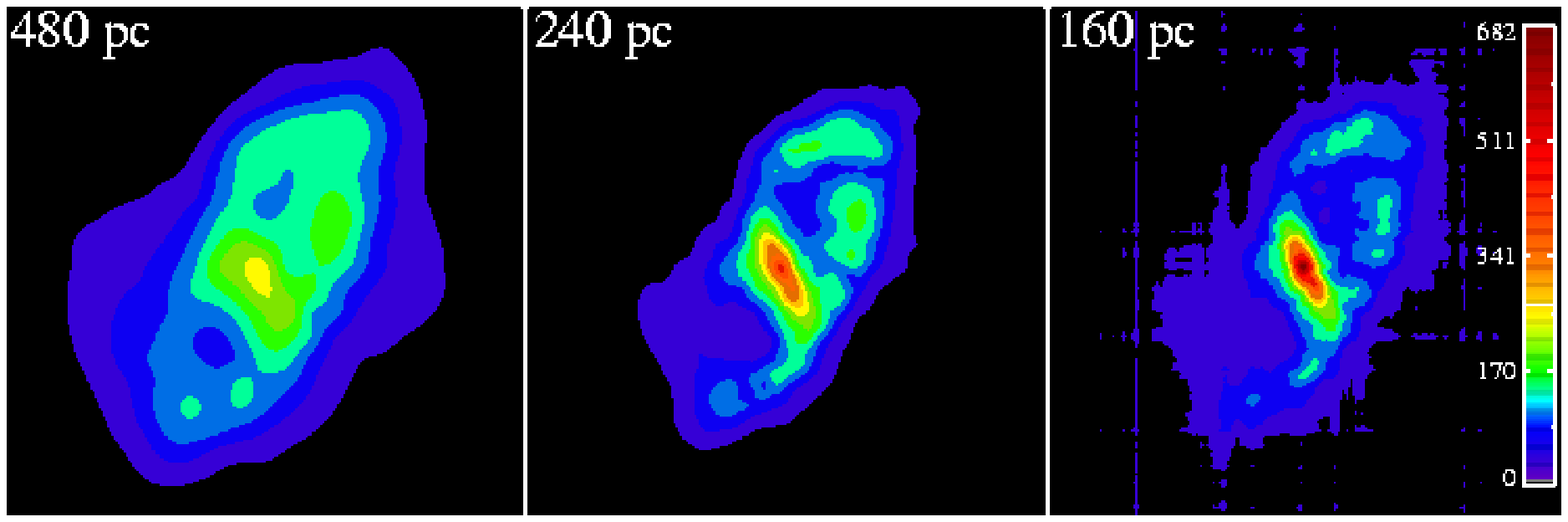}
\includegraphics[scale=0.9]{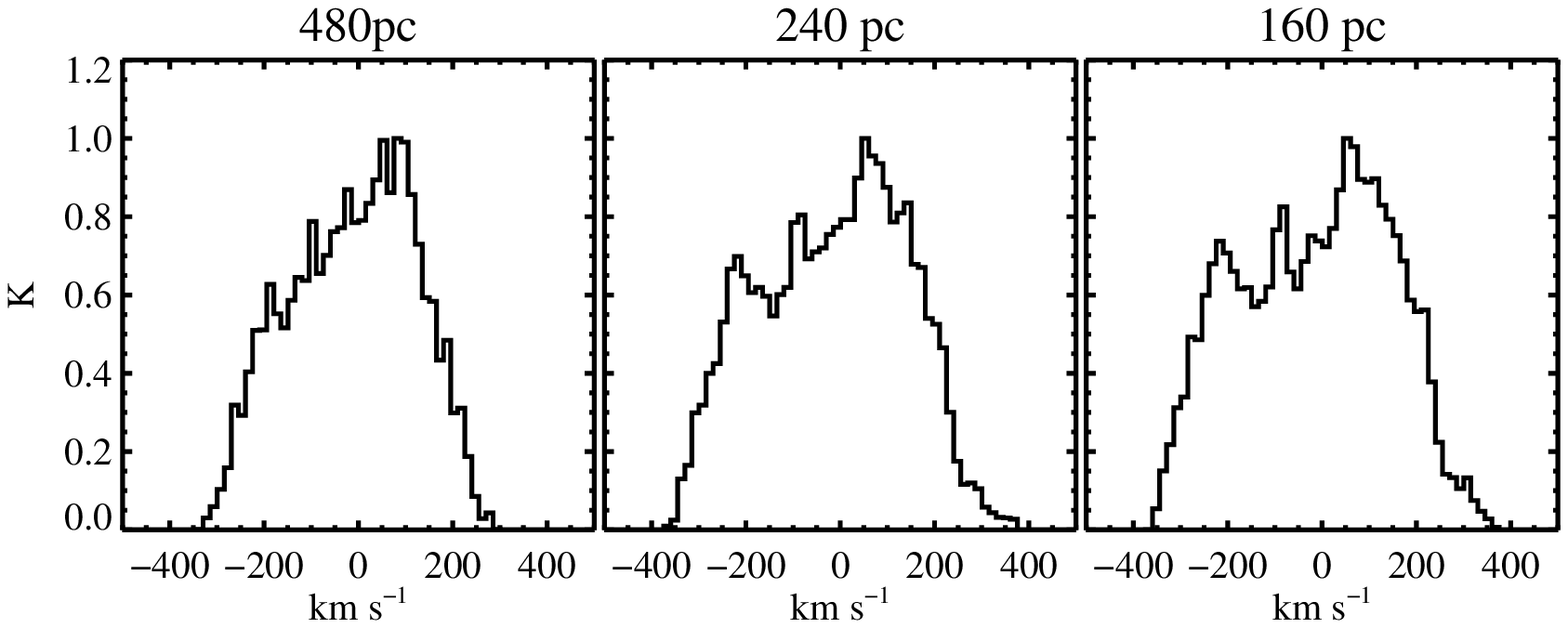}
\caption{Simulated CO (J=1-0) images and spectra for model BH at peak
of quasar phase. Images and spectra are shown for three model
resolutions. The detail in both the morphology and kinematic imprints
on unresolved emission line become apparent by model resolution of
$\sim$240 pc. \label{figure:restest_image}}
\end{figure*}

\begin{figure}
\includegraphics[scale=0.35,angle=90]{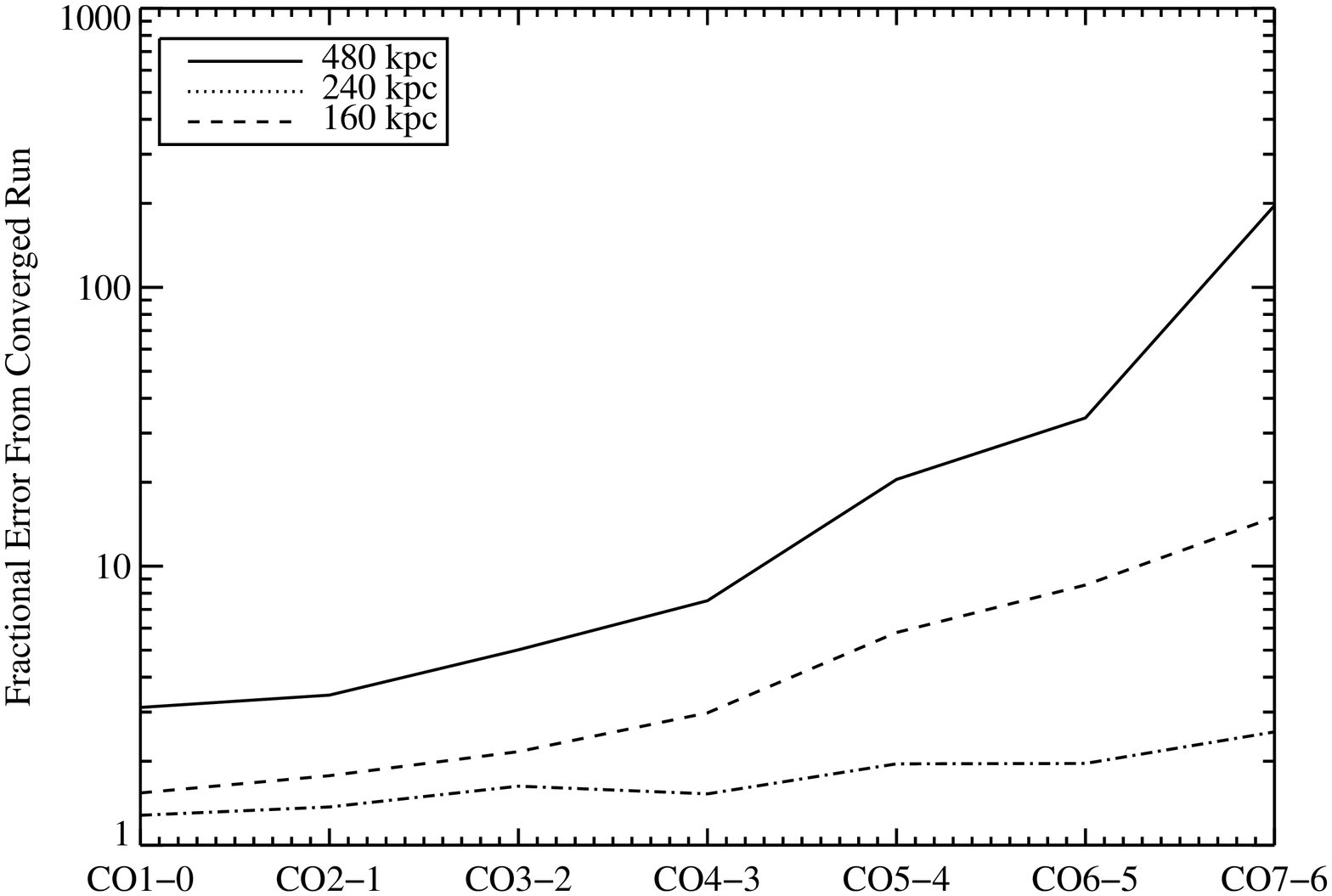}
\caption{ Ratio of total CO flux (over a variety of CO transitions)
from highest resolution model ($\sim$120 pc) to lower resolution
runs. The test snapshot is model BH during the peak of the quasar
phase. The emission from higher lying transitions may be
underestimated in cases of extremely poor spatial
resolution.\label{figure:sed_restest}}
\end{figure}

In order to test the convergence properties of the presented results,
we have taken a snapshot of merger BH (black hole winds only) at the
peak of the quasar phase, and examined the simulated CO properties at
a variety of spatial resolutions. In
Figure~\ref{figure:restest_image}, we plot the CO (J=1-0) images of
the sample galaxy at three resolutions, and the normalized CO (J=1-0)
spectra at the same resolutions. The detailed structure of the
galaxy's morphology begins to become evident at resolutions of
$\sim$240 pc, though of course shows smaller-scale features at finer
resolution.  Similarly, at 240 pc resolution the unresolved spectra
begin to show detailed kinematic signatures in the line profile that
are smoothed over in the lower spatial resolution simulations. Thus,
for both image analysis and spectral line profile analysis we utilize
simulations with $\sim$240 pc resolution in order to glean the maximum
information while minimizing the extensive computational costs
associated with galaxy-wide non-LTE radiative transfer.

Knowledge of the excitation properties of CO requires a slightly
higher spatial resolution. While the work presented here does not
nominally investigate the excitation properties of the CO molecular
gas to great detail, other investigations utilizing models similar to
these (e.g. Narayanan et al. 2007b) examine the excitation properties
in greater detail. We therefore briefly review the convergence of the
CO excitation. The excitation properties require a higher spatial
resolution for convergence, and we therefore include a higher
resolution simulation (120 pc) for comparison. In
Figure~\ref{figure:sed_restest}, we plot the ratio of the intensity of
the highest resolution run (120 pc) versus lower resolution
simulations for a series of CO transitions. The results for higher
lying transitions (CO$\ga$4) are only converged for higher spatial
resolution ($\sim$160 pc) than is typically required for the
morphology or spectral line profile studies. Therefore higher
resolution models must be accounted for in studies concentrating on
the molecular excitation (e.g. Narayanan et al. 2007b). For the few
excitation condition investigations in this work, we employ targeted
higher resolution simulations.

\subsection{Overview of A Major Merger}

\begin{figure}
\includegraphics[scale=0.525]{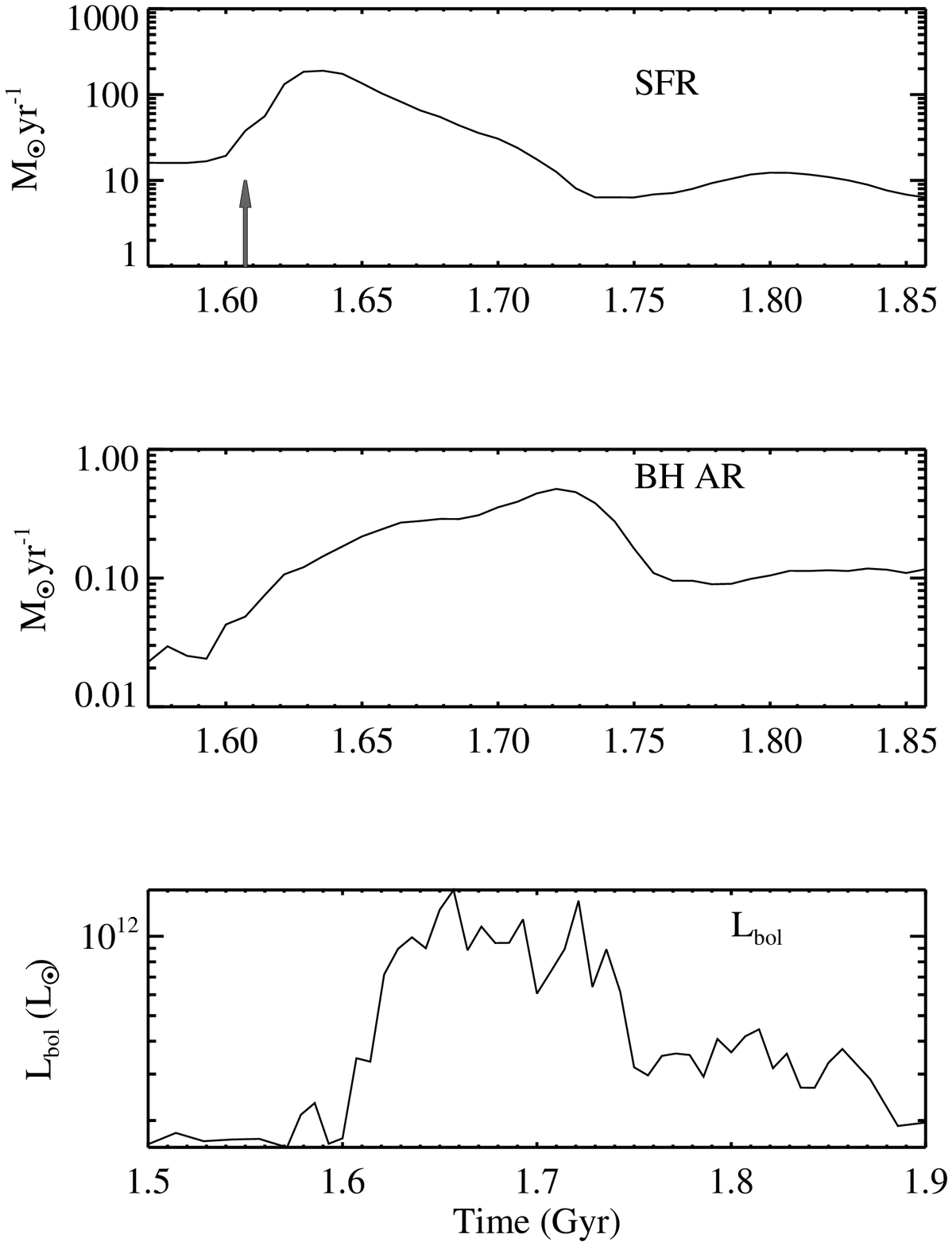}
\caption{SFR, black hole accretion rate and bolometric luminosity as a
function of time for model BH. The arrow in the top panel denotes the
point of nuclear coalescence. \label{figure:vc3vc3e_2b_bharsfr}}
\end{figure}

In the binary merger simulations presented here, after the initial
passage of the merging galaxies, gas is driven into the nuclear
regions owing to tidal torquing (Barnes \& Hernquist, 1991,
1996). This begins a major black hole growth phase. While the galaxies
approach toward and undergo final coalescence\footnote{Nuclear
Coalescence is defined for the purposes of this paper as when the
progenitor BH particles are separated by less than a smoothing length,
and are thus no longer distinguishable. BH particles are included for
all models in this paper, although they do not accrete in model
no-winds.  Coalescence could equivalently be defined as when the
stellar bulges of the progenitors are no longer distinguishable,
though in practice the choice of a definition for nuclear coalescence
is somewhat arbitrary and makes little difference in the results. The
time of coalescence is typically T$\sim$1.6 Gyr, though varies
slightly from model to model.} (hereafter defined as the in-spiral
stage) they go through a massive starburst and LIRG/ULIRG phase while
the central black hole grows as an enshrouded AGN following the
methodology developed by Springel et al. (2005a). In the models which
include AGN feedback-driven winds, after nuclear coalescence, the
thermal energy associated with the black hole accretion drives a
powerful wind into the surrounding ISM and strongly reduces the star
formation rate.  During much of this time, along several sightlines,
the object may be viewed as an optical quasar (Hopkins et al. 2005a-d;
2006a-d; 2007h).  The galaxy then proceeds to evolve passively into a
red elliptical galaxy (Springel, Di Matteo \& Hernquist, 2005b).  In
Figure~\ref{figure:vc3vc3e_2b_bharsfr}, we show as a reference the
SFR, black hole accretion rate, and bolometric luminosity of model BH
throughout the time period considered in this work.

The galaxies and quasars formed via this formulation for AGN feedback
in galaxy merger simulations have proven consistent with observed
quasar luminosity functions and lifetimes (Hopkins et al. 2006a-d,
2006a,c,d; though see Richards et al., 2006), as well as the locally
observed \magorrian relation (Di Matteo et al. 2005; Hopkins et
al. 2007a,b; Robertson et al.  2006a,c). These simulations have also
shown successes in reproducing observed X-ray, IR and CO patterns
characteristic of ULIRGs and quasars (Cox et al. 2006b Chakrabarti et
al. 2007a,b; Narayanan et al. 2006a, respectively), the bimodal galaxy
color distribution (Springel, Di Matteo \& Hernquist, 2005b; Hopkins
et al.  2006b), Seyfert galaxy luminosity functions (Hopkins \&
Hernquist, 2006), the kinematic structure of merger remnants (Cox et
al. 2006a), the SFR-CO and SFR-HCN relation in galaxies (Narayanan et
al. 2007b), and observed properties of \zsim 6 quasars (Li et
al. 2007a,b; Narayanan et al. 2007a).

\subsection{General Wind Properties in the Simulations}
\label{section:generalproperties}

\begin{figure}
\includegraphics[angle=90,scale=0.35]{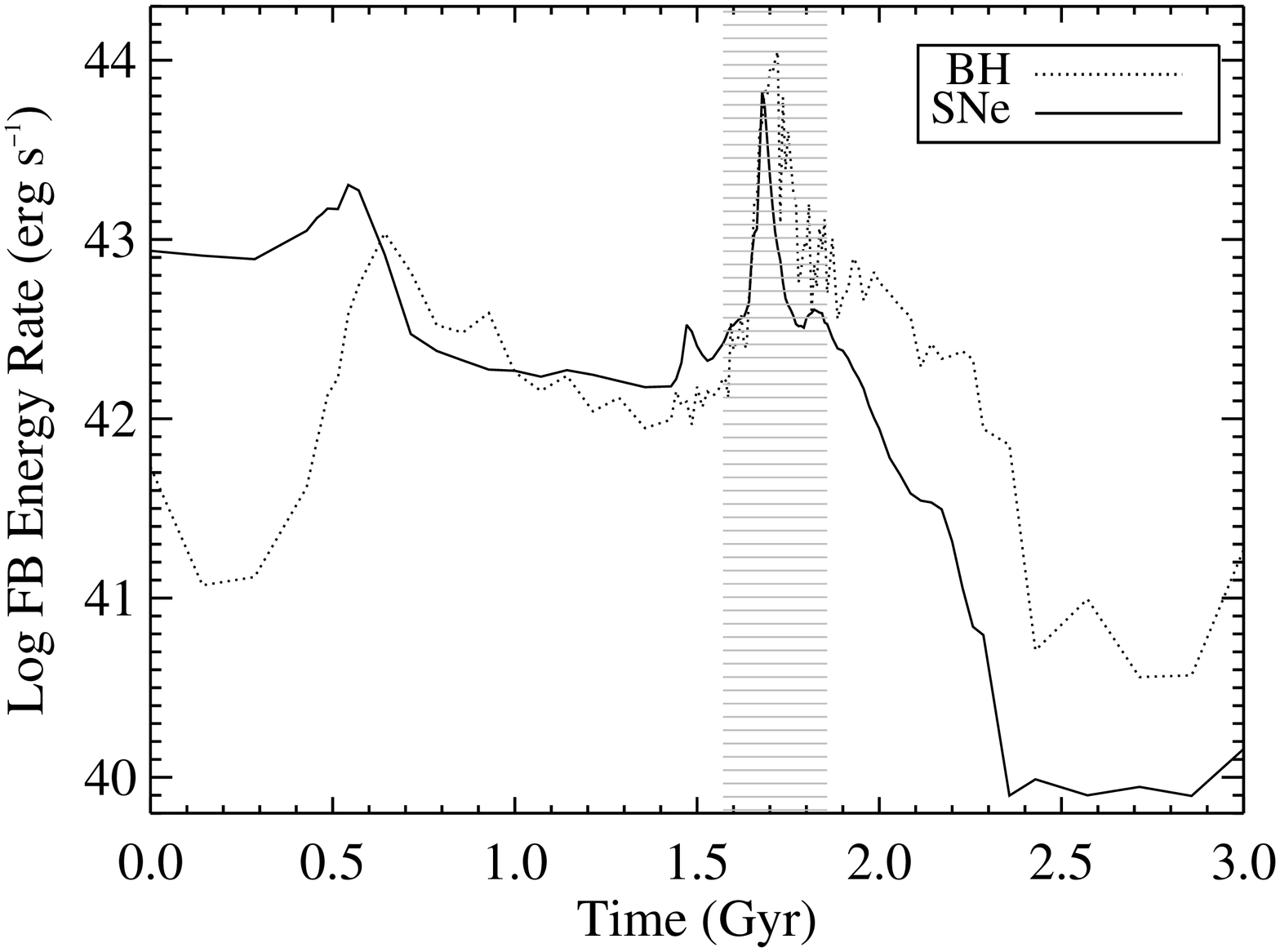}
\includegraphics[angle=90,scale=0.35]{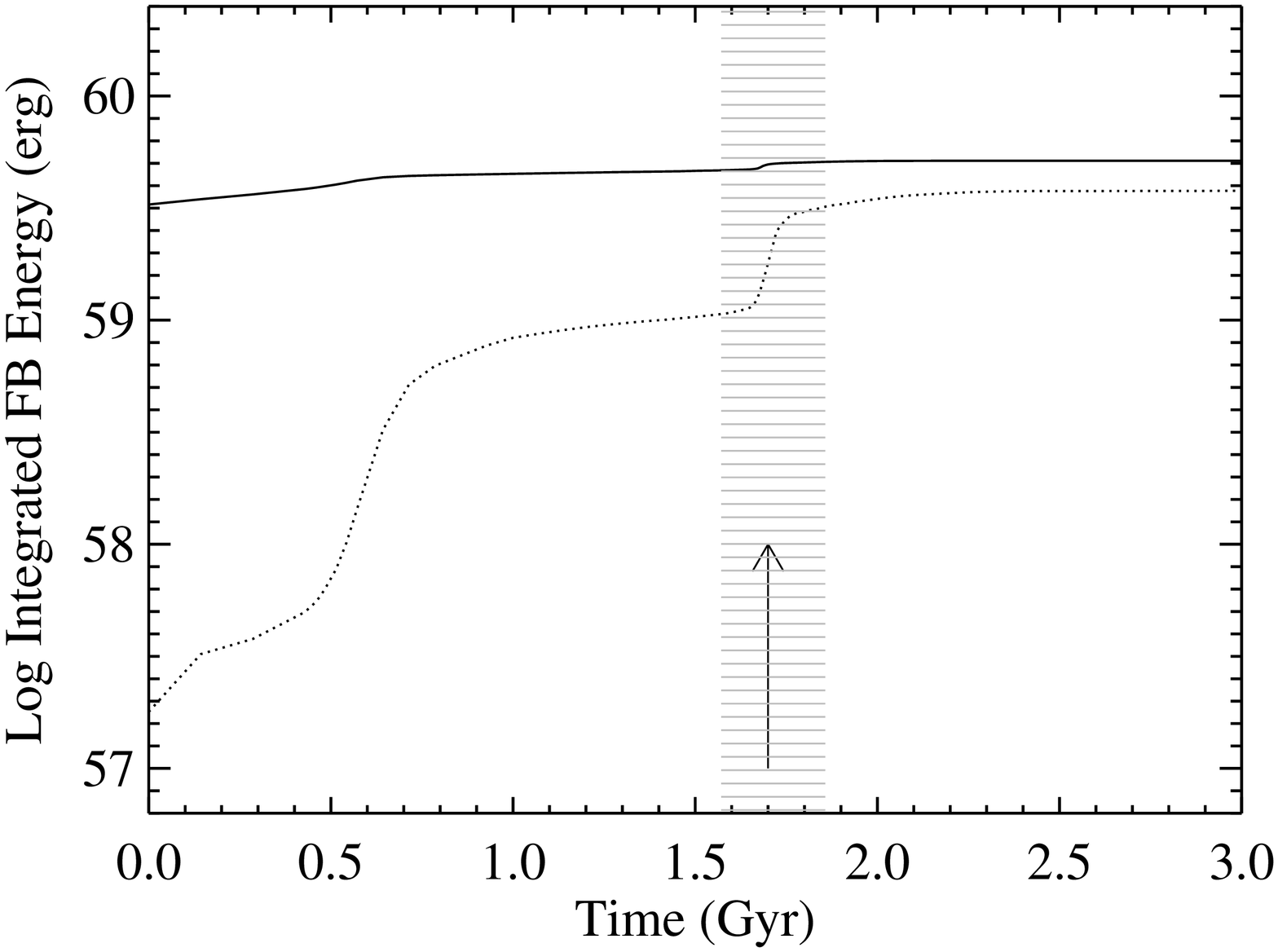}
\caption{Instantaneous feedback energy rate ({\it top}) and integrated
feedback energy ({\it bottom}) input into the ISM by black holes and
star-formation driven winds for model sbBH. The hatched region is the
$\sim$0.3 Gyr period of peak star formation and black hole activity
which will be considered for the remainder of this paper.  While the
integrated feedback from star formation and black holes is roughly
equivalent, the star formation delivers its energy in a slow and
steady manner from simulation time T=0 spread out over much of the
galaxy, in contrast to the near point explosion characteristic of the
black hole winds. The arrow in the bottom panel denotes the point of
nuclear coalescence. Model results first published in Cox et al.
(2007).\label{figure:fb_energy}}
\end{figure}

In principle, there are three mechanisms for driving winds in our
major merger simulations: via shock heated gas (Cox et al. 2004,
2007), starburst-driven winds (Cox et al., 2006c, 2007) and
AGN-feedback driven winds (Cox et al., 2007; Hopkins et al., 2005a-d;
Springel, Di Matteo \& Hernquist 2005a,b). In practice, however,
outflow rates in the simulations are dominated by winds driven by
starbursts and AGN feedback. This was explicitly shown by Cox et al.,
(2007) who found that during the active phase, the outflow rate owing
to shock heated gas was at most an order of magnitude less than that
from starbursts and/or AGN. Thus for the rest of this work, we will
focus primarily on the effects of winds from star formation and
AGN-feedback, and rely on the model without these winds (though with
shock-heated winds) as the fiducial comparative model. We note that
Table~ 1 refers to this model as 'no-winds' though in practice winds
owing to shock heated gas do exist, albeit with minimal impact on the
molecular ISM.

We next review some of the broader physical effects generated by AGN
feedback-driven and supernovae-driven winds. In a recent comprehensive
study of galactic winds in mergers utilizing (among others) identical
models to those presented here, Cox et al. (2007) analyzed the
decoupled properties of starburst and AGN feedback-driven winds in
galaxy merger simulations. We refer the reader to this work for a
thorough analysis of these effects, and highlight here the wind
properties most relevant to the discussion of this paper.

As a reference, we first show the instantaneous and integrated
feedback energies for model sbBH in Figure~\ref{figure:fb_energy}
(e.g. Equations~\ref{eq:bhenergy} and~\ref{eq:sbenergy}; model results
first published in Cox et al. 2007). We highlight the period of peak
starburst and AGN activity which we will focus on for the remainder of
this study, and denote as the ``active phase''. We focus on this
period as the models exhibit their peak wind activity during this
time.

 The first passage of the galaxies triggers a starburst event, fueling
star formation rates to $\sim$50\msunyrend, irrespective of the
presence of a central black hole. At this point, starburst-driven
winds begin to disrupt the central cold gas supplies, limiting to some
degree (dependent on the wind efficiency and speed) the magnitude of
the starburst. In the simulations presented here, during the in-spiral
stage prior to nuclear coalescence, $\sim$90\% of the star formation
(and thus $\sim$90\% of the starburst generated wind mass) occurs (Cox
et al. 2007).

The growth of the central AGN and associated feedback winds develops
along a markedly different story line. While the initial passage and
in-spiral stage of the galaxy merger fuels central black hole growth,
the majority of black hole growth occurs during the active phase
highlighted in Figure~\ref{figure:fb_energy}, typically around nuclear
coalescence.  The exact percentage of black hole growth that occurs
during the active phase is dependent on galaxy mass (Cox et al.
2007), but the generic trends are robust.

After the active phase, the total integrated feedback energy deposited
into the ISM from star formation and black holes is roughly the
same. However, the instantaneous properties are different. While the
energy input from star formation winds is spread out over the entire
in-spiral and active phase, the black hole offers a relatively short
impulse of energy. The slow buildup of the stellar mass and relatively
rapid black hole growth phase results in black holes contributing over
five times the energy input to the ISM as stars during the active
phase (Cox et al. 2007). This will have important consequences on the
relative impact of winds on CO emission from starbursts and AGN.
Throughout this work, we will focus on the rough peak of the active
phase, specifically T$\sim$1.6-1.8 Gyr.

\section{Isolated Disk Galaxy}
\label{section:spiral}


\begin{figure}
\includegraphics[scale=.5]{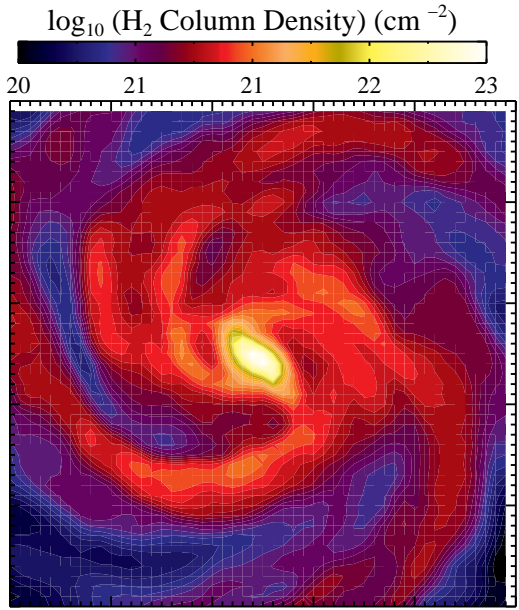}
\caption{\htwo \ column density for fiducial progenitor disk galaxy (CO
contours shown in Figure~\ref{figure:spiral}). The panel is 12 kpc on
a side, and the scale is on top.\label{figure:cdens}}
\end{figure}

\begin{figure*}
\plotone{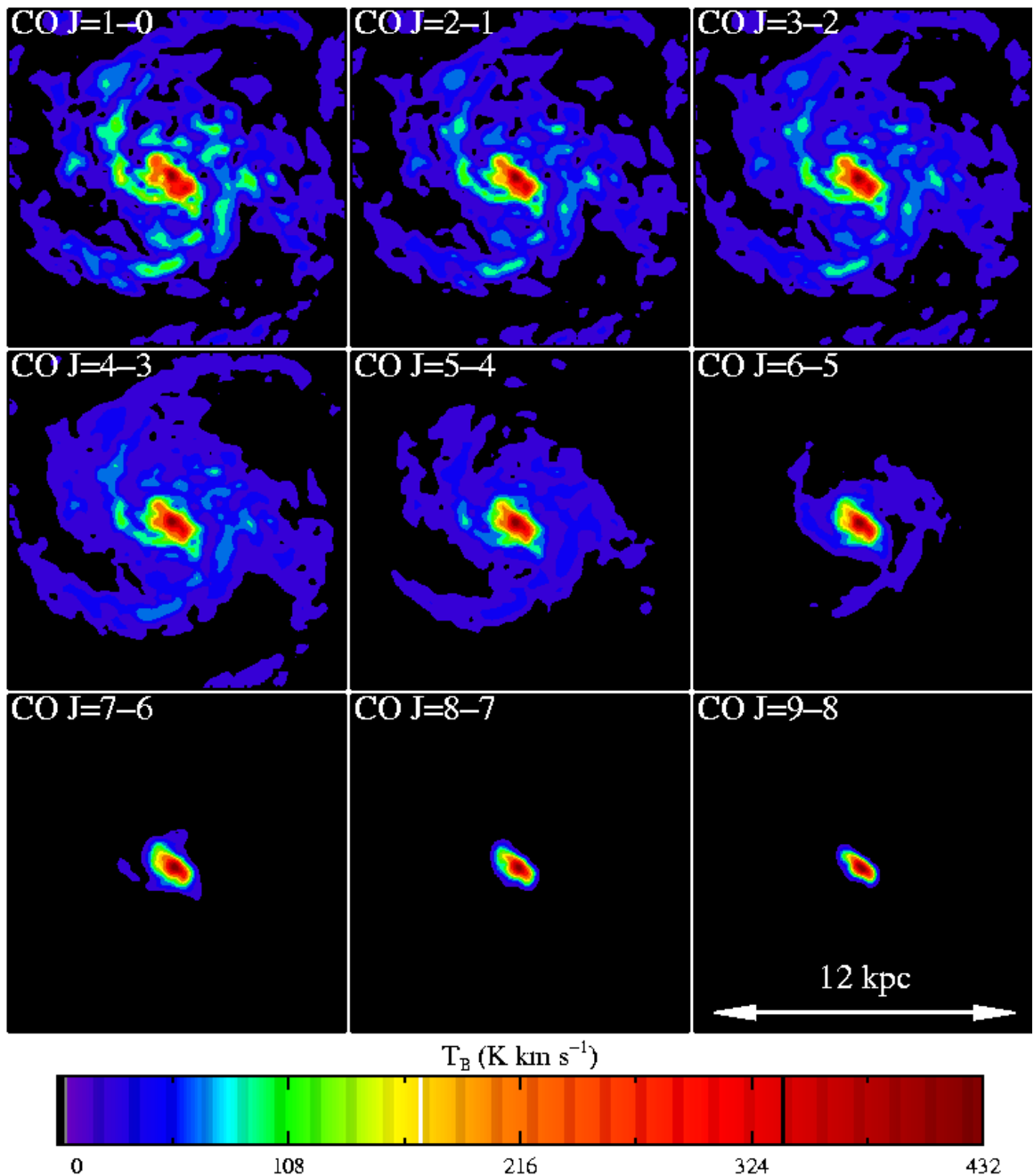}
\caption{Simulated CO (J=1-0) through (J=9-8) emission contours for
progenitor disk galaxy. While lower CO transitions trace the bulk of
the molecular gas, higher lying transitions with relatively high
critical densities probe only the nuclear star forming regions. Panels
are 12 kpc on a side, and scale on bottom is in units of
K-\kmsend.\label{figure:spiral}}
\end{figure*}

\begin{figure}
\includegraphics[scale=0.35,angle=90]{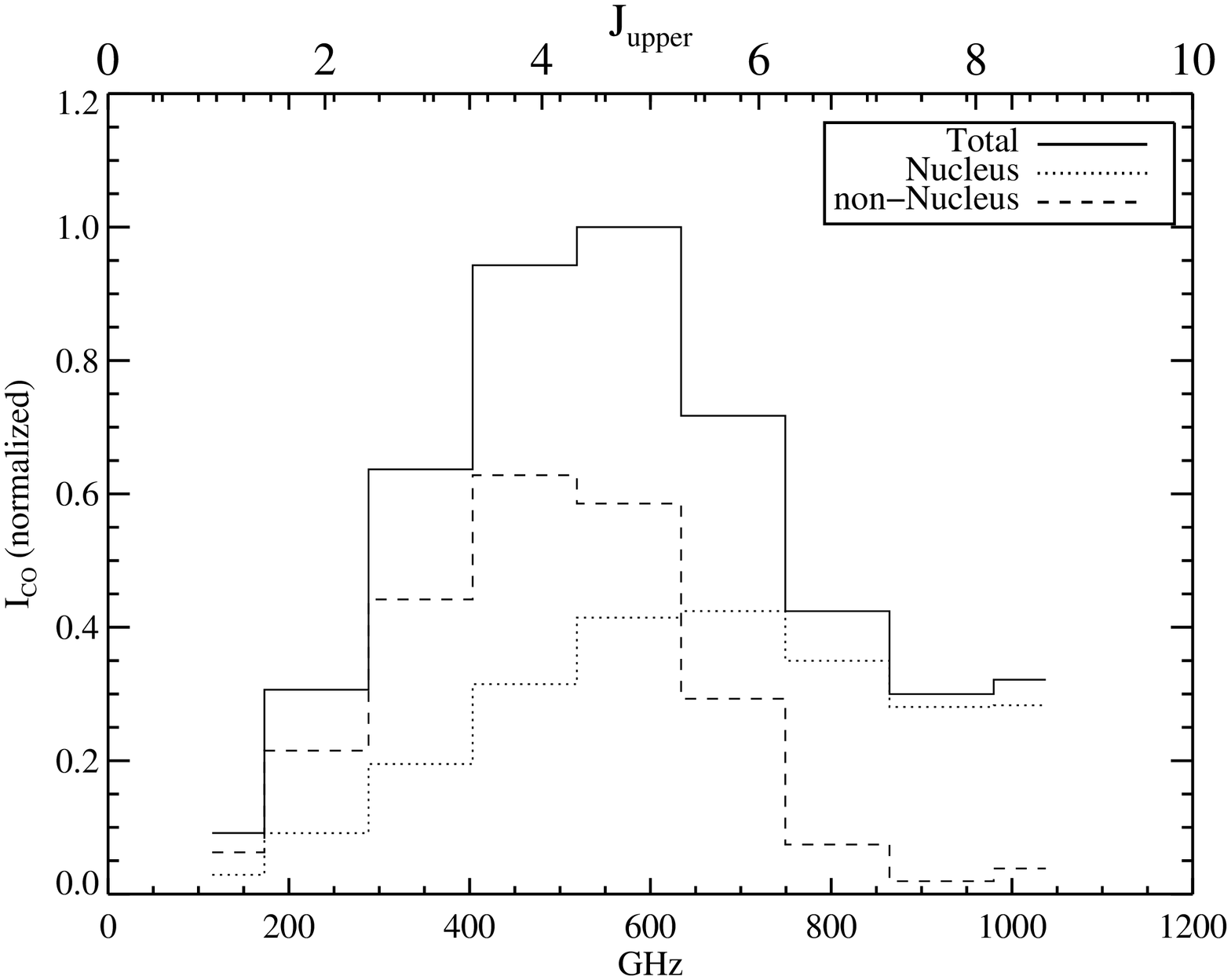}
\caption{Normalized CO SED of representative progenitor disk
galaxy. The CO SED is decomposed into contribution from the nuclear
(central kpc) region and the rest of the galaxy.  The CO SED shows the
relative flux density for different CO transitions, and plots the
frequency of the transition on the bottom axis and upper J state on
the top axis. The peak in the total CO SED as well as that just from
the starbursting nucleus is consistent with results derived from local
starburst galaxies, supporting our methodology for deriving CO
excitation patterns in galaxy simulations (Bradford et al. 2005;
Wei\ss \ et al. 2005b).\label{figure:cosed}}

\end{figure}

To illustrate the generic properties of our methods, we have allowed a
fiducial 40\% gas fraction progenitor disk galaxy to evolve as
outlined in Springel et al. (2005a). For reference, in
Figure~\ref{figure:cdens}, we plot the \htwo \ column density map
through the model disk galaxy. In Figure~\ref{figure:spiral} we
present the CO (J=1-0) through (J=10-9) intensity contours for
comparison with observations where the intensity units are K$\cdot$km
s$^{-1}$ where the temperature unit is a Rayleigh-Jeans temperature.

 The CO (J=1-0) emission traces the star forming molecular gas through
the spiral arms and becomes more intense near the starbursting nuclear
region. Additionally visible are individual concentrations of cold gas
and star forming regions emitting as discrete pockets of CO (J=1-0)
emission throughout the spiral arms. Higher density CO transitions
probe gas of a higher critical density. Thus the higher-lying
transitions probe the higher density peaks, and become more centrally
concentrated, following the density distribution in the galaxy.

At first glance, the emission contours from e.g. the CO (J=1-0) image
appear to peak at higher temperatures ($T_B$) than are typically
recorded from observations of local star forming spirals. This effect
owes to the spatial resolution of the radiative transfer models
($\sim$250 pc). When simulating more practical observing conditions,
the results from our modeling are quite comparable to observations of
local universe spirals. As an example, we compare our models to
50$\arcsec$ observations by Scoville \& Young (1983) of CO (J=1-0)
emission from M51. When our model disk galaxy was scaled to be at the
distance of M51 and the CO (J=1-0) emission convolved to the same
resolution as the Scoville \& Young observations, the integrated
intensity in our model was comparable to the $\sim$ 60 K-\kms observed
in M51 (Scoville \& Young, 1983).

In Figure~\ref{figure:cosed} we present the normalized CO line
spectral energy distribution (SED) of the model face-on disk galaxy.
The CO SED represents the total integrated flux density emitted from
each line and serves as a diagnostic for the excitation conditions in
the molecular gas. These are plotted as the CO integrated flux density
versus the upper rotational quantum number of the transition. CO SEDs
are particularly useful in studies of unresolved high-$z$ sources
(e.g.  Narayanan et al. 2007a; Wei\ss \ et al. 2005a, 2007).  We have
also decomposed the CO SED by plotting the contribution to the flux
from the starbursting nucleus (central kpc) and from the rest of the
galaxy.

The CO flux density rises up to the CO (J=5-4) transition, and then
drops toward higher excitation lines.  The intensity from the highly
excited component of the CO gas (e.g. J>6) comes primarily from the
warm and dense conditions in the nuclear starburst regions, though
some contribution from subthermally excited gas in the outer regions
of the galaxy is visible (Narayanan et al. 2007b).  This is in good
agreement with observations of local starbursts. For example, in a CO
SED decomposition of M82, Wei\ss \ et al. (2005b) found that the
central regions peak at the CO (J=6-5) transition whereas the total
SED peaks near the CO (J=4-3) transition. Similarly, the CO line SED
from the nuclear region of NGC 253 peaks near J=7 or J=8 (Bayet et
al. 2004; Bradford et al. 2003, 2005)

\section{CO Morphology in Mergers with Winds}
\label{section:morphology}
Our approach in combining non-LTE radiative transfer calculations with
SPH simulations allows us to uniquely explore, for the first time, a
variety of the decoupled effects of starburst and AGN feedback-driven
winds on the molecular gas emission from galaxy mergers.  In
particular, a powerful tool for understanding the spatial distribution
of molecular gas in local mergers is through high resolution
velocity-integrated intensity contour maps. These can serve to
describe the morphology and kinematics of the star-forming molecular
gas, as well as characterize the nature of star-forming regions
through observations of multiple excitation levels (e.g. Iono et
al. 2004). Indeed, such maps have revealed, for example, the compact
nature of molecular emission in mergers (Bryant \& Scoville, 1999),
the double-nucleus in the central regions of Arp 220 (Sakamoto et
al. 1999), and the large scale distribution of atomic and molecular
gas in colliding galaxy systems (Iono, Yun \& Ho, 2005). In this
section, we build on the current understanding of CO morphologies in
interacting systems by describing the response of the CO morphology to
starburst and AGN feedback-driven winds. We first discuss general
properties of the CO morphology in mergers. We then continue with a
discussion of signatures of galactic winds apparent in CO emission
maps via outflows detectable in the CO morphology. We conclude the
section with a brief discussion of secondary effects of galactic winds
in affecting the spatial extent of CO gas.

\subsection{General Properties of CO Morphology}
\label{section:generalmorph}

\thispagestyle{empty}
\setlength{\voffset}{-18mm}
\begin{figure*}
\plotone{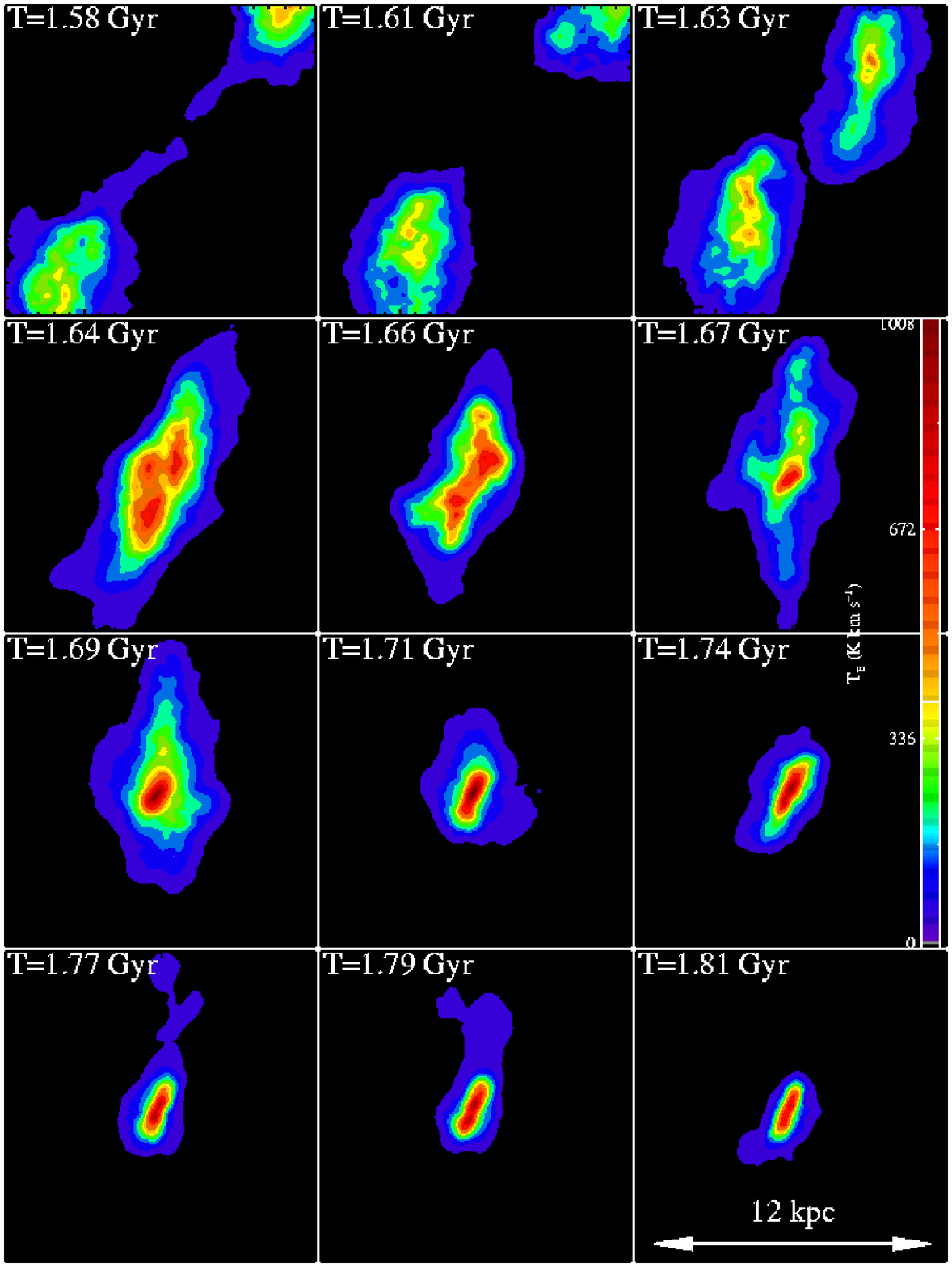}
\caption{Contours of velocity-integrated CO (J=1-0) emission at
various snapshots through evolution of merger model sbBH, with both
starburst and AGN feedback-driven winds. The results discussed here
are generic for all wind models. Upon nuclear coalescence (T=1.7 Gyr)
a burst of star formation drives the CO luminosity to near peak
values. The CO luminosity then fades as star formation consumes the
available molecular gas. Units are K-\kms (scale at bottom), and each
panel is 12 kpc on a side.
  \label{figure:sb10BH_morph}
}
\end{figure*}

\setlength{\voffset}{0mm}

In Figure~\ref{figure:sb10BH_morph}, we show the evolution of the CO
(J=1-0) emission during a major merger of two disk galaxies simulated
in run sbBH (model with starburst and black hole winds).  The
properties discussed in this subsection
(\S~\ref{section:generalmorph}) are generic to each of our models,
regardless of wind parameter choice. The images in
Figure~\ref{figure:sb10BH_morph} start after the first passage of the
galaxies, when they are approaching for their final coalescence. The
tidal interaction of the first passing of the galaxies causes
`streamer' gas to emit in the intergalactic space between the two
galaxies. The CO morphologies are disturbed by the first passage
itself, and the CO emission is extended in nature until the nuclei
have fully merged. The first passage of the galaxies induces a burst
of star formation, peaking in the nuclear regions of each galaxy. This
is reflected in the luminous CO emission at the center of each galaxy.

Because the CO (J=1-0) can be thermalized at relatively low critical
densities, and lies only $\sim$5 K above ground, it serves as a good
tracer for total molecular gas content. Both diffuse envelopes of
clouds as well as dense cores can emit CO (J=1-0) flux.  Conversely,
emission from higher lying lines of CO, and high dipole moment
molecules (e.g. HCO+, HCN, CS) arises primarily from dense cloud
cores. Thus, higher lying CO lines (e.g. CO J=3-2) from the merger
broadly follow the distribution of the CO (J=1-0) line, but remain
more compact in nature, generally emitting only from the dense
starbursting nucleus. This was explicitly seen in the disk galaxy in
Figure~\ref{figure:spiral} as well.

 The nuclear CO intensity peaks soon after the progenitor galaxies lose
individual identity, and there is only one surface brightness peak.
During this time, stars formed in the merger-induced starburst heat up
the molecular gas such that collisions readily excite the CO in the
warm (T$\sim$30 K) and dense ($n\sim$10$^{5}$\cmthree) gas. The
columns are still quite large through the nucleus as the starburst has
not had time to consume the available molecular gas.

While the nuclear CO intensity peaks near coalescence, the globally
averaged velocity-integrated CO intensity across all transitions drops
through the evolution of the merger. This can be attributed to global
consumption of cold molecular gas by star formation over time.  This
result is consistent with observations by Gao \& Solomon (1999) who
found from CO (J=1-0) emission that the total molecular gas content of
a merger decreases as a function of projected nuclear separation in
galaxy mergers.

\subsection{Molecular Outflows}
\label{section:outflowmaps}

Starburst and AGN feedback-driven winds can leave an imprint on the
molecular line emission via large scale molecular outflows seen in the
CO morphology.  These winds are seen to entrain large masses of
molecular gas in outflows which appear prominently as a secondary
surface brightness peak in velocity-integrated emission contour maps
(Narayanan et al.  2006a).

In this section, we discuss the properties of the CO morphology in the
context of these outflows.  The outflows discussed here are only those
which are large enough to be visible via CO (J=1-0) emission maps.
For the purposes of understanding general trends of detectable
outflows in this section, we will find it convenient to define a
detectable outflow as any CO emission peak at least 50\% of the
nuclear emission maximum, with a projected distance at least 1.5 kpc
away from the galactic nucleus. The molecular outflows are defined
only after nuclear coalescence, and thus also follow the requirement
that the CO emission peak be separate from the stellar surface density
peak.

Before continuing, we note that the simulations chosen here are not
extreme in any way (Figure~\ref{figure:vc3vc3e_2b_bharsfr}). The peak
B-band luminosity during the quasar phase is
$\sim$10$^{11}$\lsun. Similarly, the peak IR luminosity is
$\sim$10$^{12}$\lsun, allowing the galaxy to be just visible as a
ULIRG. More massive mergers (e.g. Cox et al. 2007), and, as will be
discussed in \S~\ref{section:morphorientation}, more coplanar mergers
will drive stronger winds, leaving the simulations in this section
representative of lower limits.

\subsubsection{Outflow General Properties}
\label{section:outflowmapgeneral}
Here, we discuss the general properties of both starburst and AGN
feedback-driven outflows which appear in CO emission maps. While there
are differences in some specific properties of the outflows based on
the type or strength of the driving wind, we defer those comparisons
to later subsections and focus this subsection on the outflow
properties shared by all the simulations presented here.

\begin{figure*}
\plotone{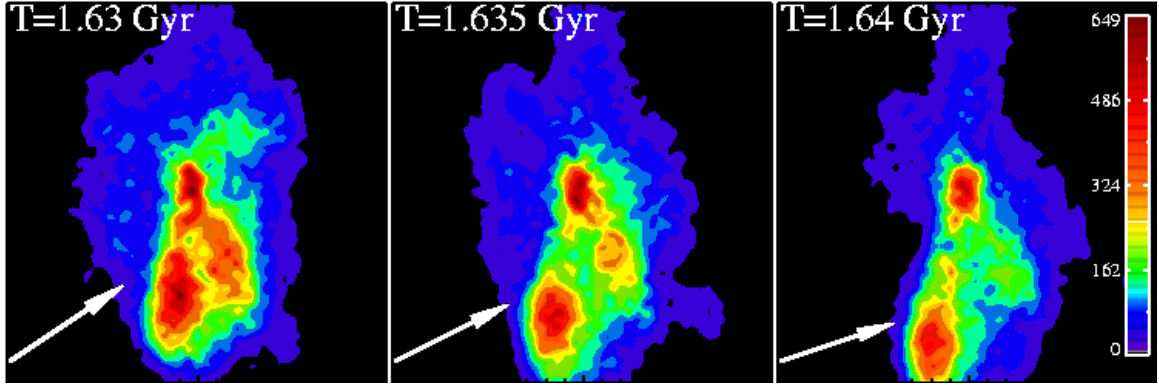}
\caption{Velocity-integrated CO (J=1-0) emission map of detectable
molecular outflows seen in the model with black hole winds only (see
also Narayanan et al. 2006a). Thermally driven winds associated with
black hole growth can entrain large masses ($\sim 10^8-10^9$\msun) of
molecular gas, resulting in directly detectable outflows.  For
reference, nuclear coalescence is at T=1.6 Gyr. The units are K-\kms
(scale on right), and the panels are 12 kpc each.
\label{figure:vc3vc3e_outflow}}
\end{figure*}

In Figure~\ref{figure:vc3vc3e_outflow}, we show an example of
molecular gas entrained in the outflow from model BH (AGN winds only;
outflow first described in Narayanan et al. 2006a).  The cold gas in
the outflow is not completely ablated by starburst and AGN
heating. While the CO emission from cold gas in the outflows closest
to the nucleus is faint owing to lower columns of cold molecular gas,
clouds which remain more deeply embedded in the outflow continue to
exhibit active star formation. During short ($\sim$5 Myr) time
intervals, the CO (J=1-0) velocity-integrated intensity from galactic
outflows can be seen to be comparable (within a factor of $\sim$2) to
the galaxy's nuclear emission.

The detectable outflows typically correspond to entrainments of mass
$\sim$10$^{8}$-10$^{9}$\msunend, with peak \htwo \ column of
$\sim$10$^{22}$-10$^{23}$\cmtwo \ (within the constraints of our
definition of a detectable outflow, \S~\ref{section:outflowmaps}).
The more massive outflows in this mass range tend to be seen closer to
the point of nuclear coalescence, with a trend toward decreasing
outflow mass as the merger progresses. This owes to more loosely bound
gas near the time of coalescence, as well as the feedback energy rate
being near its peak (Figure~\ref{figure:fb_energy}). These outflow
masses are typically 5-15\% of the total molecular gas mass.  In the
mergers e, sb, and sbBH, with random disk orientation
angles\footnote{Though note that only one random orientation is
considered in this paper.}, the outflows do not occur in any preferred
direction. As we will discuss later, this is not the case for the
coplanar model.

\begin{figure*}
\plotone{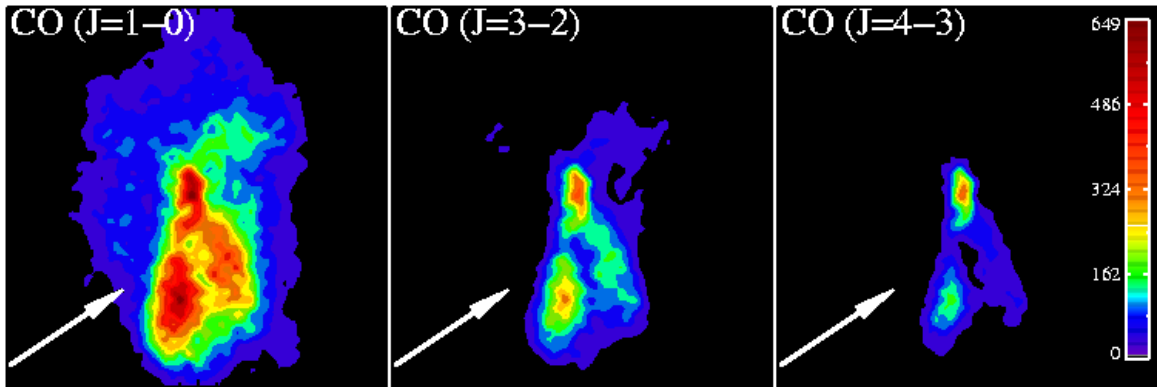}
\caption{First panel of Figure~\ref{figure:vc3vc3e_outflow} in
commonly observed CO transitions J=1-0, J=3-2, and J=4-3. The massive
and dense outflows entrain numerous GMCs, and continue to actively
form stars. Dense cores in entrained GMCs emit strongly in
higher-lying CO transitions, and can be imaged in J=3-2 and J=4-3, at
submillimeter wavelengths.  The outflow lifetime typically drops with
transition as the outflows become more diffuse further from the
nuclear regions (see text). For reference, nuclear coalescence is at
T=1.6 Gyr. The intensity scale in K-\kms, and the scale is on the
right. Each panel is 12 kpc on a
side. \label{figure:vc3vc3e_outflow_excitation}}
\end{figure*}

The entrained clouds can be relatively excited, allowing outflows to
be imageable at higher-lying CO transitions (e.g. J$\ga$4). In
Figure~\ref{figure:vc3vc3e_outflow_excitation}, we show the outflow
from the first panel of Figure~\ref{figure:vc3vc3e_outflow} in CO
transitions J=1-0, J=3-2 and J=4-3. The latter two transitions have a
relatively high critical density of $\ga$10$^4$ \cmthree meaning they
are typically thermalized only in the dense cores of GMCs, and may
rely on radiative pumping for excitation in other parts of the
ISM. The warm and dense gas heated by active star formation in the
deeply embedded clouds in the outflows collisionally excite the CO gas
into higher levels. In fact, the CO flux density in the outflow peaks
at J=4, consistent with actively star-forming gas and multi-line
measurements of excitation patterns in starburst-driven outflow of M82
(Walter, Wei\ss \ \& Scoville, 2002). It should be noted that the
relative visible lifetimes of outflows in higher-lying transitions may
be much smaller than at CO (J=1-0). For example, the outflow presented
in Figure~\ref{figure:vc3vc3e_outflow} appears as a unique secondary
surface brightness peak for $\sim$35 Myr at CO (J=1-0). In the CO
(J=3-2) transition, the same outflow is distinguishable from the
background for only $\sim$20 Myr.

\subsubsection{Relative Role of AGN and Starbursts in Powering Detectable CO Outflows}
\label{section:relativestaragnmorph}

\thispagestyle{empty} \setlength{\voffset}{-22mm} \epsscale{.85}
\begin{figure*}
\plotone{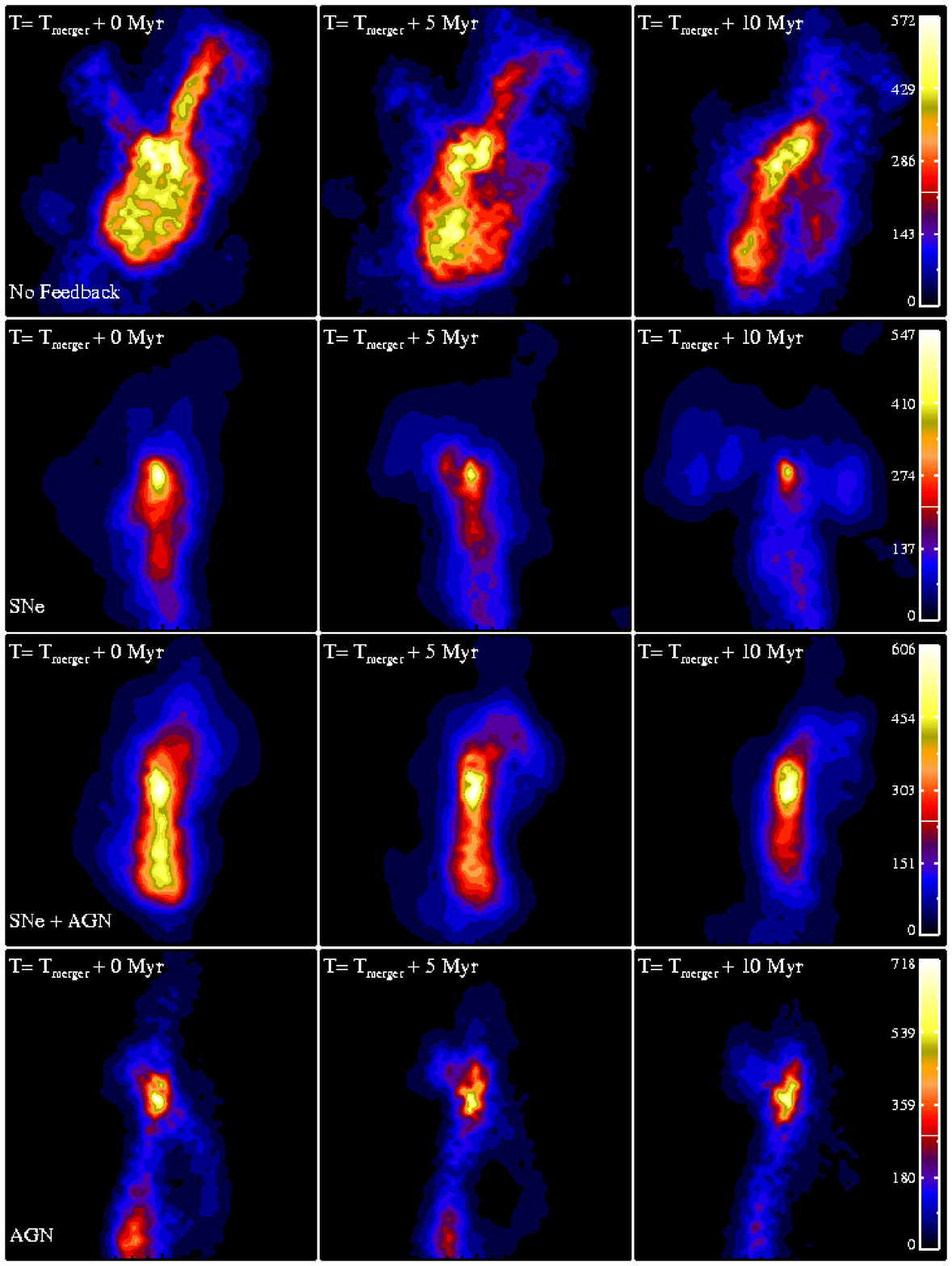}
\epsscale{1}
\caption{CO (J=1-0) emission contours in snapshots of models BH,
no-winds, sb and sbBH (thus varying only wind models) near nuclear
coalescence. The first snapshot in each row is the point of nuclear
coalescence, and the timestamps at the top left refer to the length of
time after coalescence. Loosely bound gas soon after the merger is
visible in the nuclear regions of the model without winds (no-winds,
top row). As the strength of the winds increases (second through
fourth rows), this gas is entrained and can be imaged in outflow. The
wind with the strongest instantaneous impact is the AGN-only
simulation (model BH) as the presence of starburst winds can slow
accretion onto the central black hole. Thus, the outflows in the
AGN-only wind model are the most dramatic. Each panel is 12 kpc, and
the contours are in units of K-\kms. For clarity in the images, we
have plotted each model on its own color scale, though note that the
scales are roughly similar (scale on the right of each row).
\label{figure:outflows_differentmodels}}
\end{figure*}

\setlength{\voffset}{0mm}

As an illustrative point demonstrating the relative contributions of
starburst and AGN winds in driving detectable molecular outflows, in
Figure~\ref{figure:outflows_differentmodels} we plot three snapshots
during the period of nuclear coalescence for models no-winds, BH, sb,
and sbBH (thus varying only the form of winds included, and not
geometry of the merging galaxies). The wind-dependent morphological
differences in the models are typically minimal near the point of
nuclear coalescence, and thus the outflows imaged at this time serve
as a reasonable comparison for the capacity of different wind models
to drive outflows. Here, loosely bound gas near the inner regions is
seen to be entrained in nuclear galactic winds. For reference, these
outflows are the same as those shown in
Figures~\ref{figure:vc3vc3e_outflow} and
~\ref{figure:vc3vc3e_outflow_excitation}.

Model BH is able to drive strong outflows (4th row,
Figure~\ref{figure:outflows_differentmodels}) with properties
described generally in \S~\ref{section:outflowmapgeneral}. The black
hole feedback is tightly associated with the accretion rate, and thus
the AGN wind is produced over a very short period of time and peaks
soon after nuclear coalescence.  Relative to the evolution of the
galaxy, the AGN wind generically acts like a point explosion, and the
bulk of the energy is imparted to the ISM nearly instantaneously.
Indeed, Hopkins et al. (2006c) and Hopkins \& Hernquist (2006) have
shown explicitly that these AGN-driven winds are well-characterized by
generalized Sedov-Taylor solutions for point explosions.  The outflows
in model BH are primarily seen soon after the major merger, when the
gas is still highly dynamical and loosely bound. They are also seen
near the height of quasar activity, when the AGN feedback-driven winds
are strongest.

The starburst driven winds are also able to power outflows (2nd row,
Figure~\ref{figure:outflows_differentmodels}), albeit less strongly
than in the AGN-wind models; consequently, these outflows are
detectable via emission mapping for shorter periods of time. The
typical lifetime of an outflow such that it is detectable in CO
(J=1-0) in the sb simulation (with only starburst winds) is
$\sim$20-25 Myr (compared with $\sim$35 Myr in the AGN only
model). On average, 3-5$\times$10$^{8}$\msun are entrained, although
outflows as small as $\sim$10$^{7}$\msun are seen to occur, consistent
with observations of starburst driven outflows by Sakamoto et
al. (2006) and Walter et al. (2002). It should be noted that these
values are derived within the constraints of the parameters chosen for
our starburst winds ($\eta$=0.5, $v_w$=837 \kms;
\S~\ref{section:numericalmethods}). In principle, a more efficient
wind might increase the mass outflow rate, but, as discussed in Cox et
al. (2007), such parameter choices were found to quench merger-induced
starbursts, and are thus probably not representative of e.g. ULIRGs.

The relative weakness of starburst-driven outflows compared to the AGN
only model is a direct consequence of the lower peak energy input from
the starburst-driven winds, and hence lower power.  During the active
phase, the feedback from black holes inputs more than five times the
energy as that imparted by star formation. This can be attributed to
two causes. First, black hole growth is far more efficient at
converting accreted mass into thermal energy than star formation. For
example, by utilizing Equations~(\ref{eq:bhenergy})
and~(\ref{eq:sbenergy}), Cox et al.  (2007) noted that the relative
efficiency of black holes to star formation in converting gas mass to
energy deposited into the ISM was of order $\sim$640 (see, also, Lidz
et al. 2007).  Second, during the active phase, the relative black
hole growth rate is much larger than the concomitant SFR. During this
time, while the BH grows by a factor of $\sim$4, only $\sim$8\% of the
final stellar mass is formed (Cox et al. 2007). In contrast,
$\sim$90\% of the stellar mass (and consequently, stellar winds) is
produced during the in-spiral stage, prior to the active phase.

The effect of including both starburst and AGN winds is demonstrated
in the third row of Figure~\ref{figure:outflows_differentmodels}. The
outflows produced are more massive and longer lived than the
starburst-only wind models, but less so than the AGN-only wind
models. The greater capacity to drive outflows over the starburst
winds only model (sb) can be attributed to the extra energy input
given by the accreting black hole(s). However, the outflows in
Figure~\ref{figure:outflows_differentmodels} are seen to be less
powerful than those in the black hole only model. This effect owes to
the starburst winds that occur prior to nuclear coalescence. As shown
explicitly by Cox et al. (2007), starburst winds lower the amount of
rotationally supported gas early during the in-spiral stage, which
limits the amount of gas available for immediate funneling into the
central nucleus via tidal torquing (Hernquist, 1989; Mihos \&
Hernquist, 1994b, 1996). This gas does eventually rain into the
nuclear potential, fueling black hole growth and producing a black
hole with the same mass as in the AGN wind only model (BH) and a
galaxy that lies on the M$_{\rm BH}$-$\sigma_{\rm v}$ line (Cox et
al. 2007; Di Matteo, Springel \& Hernquist, 2005; Robertson et al.
2006a), as well as the black hole fundamental plane (Hopkins et
al. 2007a,b).  However, the black hole growth is integrated over
significantly longer time, and thus the impact of the associated winds
on molecular gas emission is diminished from the AGN-only model.

\begin{figure}
\includegraphics[angle=90,scale=0.35]{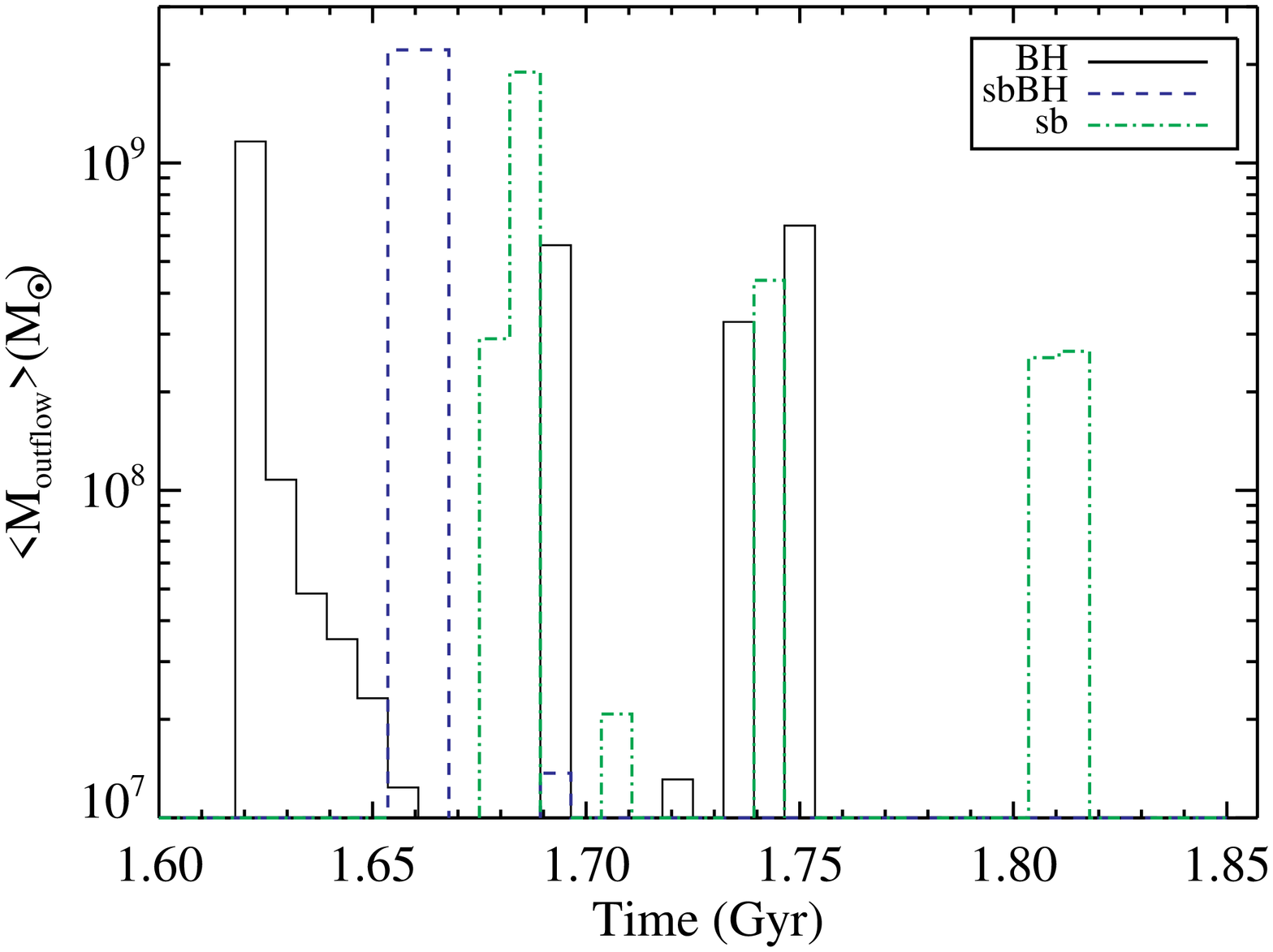}
\caption{Masses of detectable outflows as a function of time for model
BH, averaged over three orthogonal sightlines for models BH, sb and
sbBH. The masses of a typical detectable outflow are comparable
between models. The AGN-only model drives outflows preferentially near
nuclear coalescence, and for longer integrated periods of
time. Comparatively, the simulations with starburst winds generally
drive outflows for shorter periods. For reference, the times of
nuclear coalescence are T=1.6 Gyr for model BH and T=1.7 Gyr for
models sbBH and sb.
\label{figure:imageable_clumps_mass}}
\end{figure}

The relative effects of AGN feedback and starburst-driven winds on
powering molecular outflows can be summarized in
Figure~\ref{figure:imageable_clumps_mass}, where we plot the mass of
detectable outflows as a function of time for models BH, sb and
sbBH. The masses of gas each model is able to drive in an outflow is
comparable to within a factor of $\sim$2-3. AGN feedback-driven winds
drive outflows clustered primarily around nuclear coalescence and the
peak period of black hole accretion. In comparison, the models with
starburst driven winds (sb and sbBH) power detectable outflows for
relatively shorter periods.

\subsubsection{Merger Orbit Dependence}
\label{section:morphorientation}

The strength of outflows in merger models depends on the initial
orientation of the merging galaxies. In models where the disks are
aligned, strong tidal torquing can cause a rapid inflow of gas thus
providing a sudden burst of AGN feedback energy injected into the
ISM. Alternatively, geometries such as those in the BH and no-winds
models by comparison experience less tidal torquing on the gas,
causing the major black hole accretion phase to be longer lived. In
order to investigate the dependence of CO outflows on merger orbit, we
have run simulation co-BH (Table~\ref{table:ICs}) in which the
progenitor galaxies are coplanar with the orbital plane
($\phi_1$=$\theta_1$=$\phi_2$=$\theta_2$=0). The stronger tidal
torquing in the coplanar configuration results in a more efficient
blow-out of material. The \htwo \ gas in the galaxy remains confined
in a thin configuration in the plane of the merger, consistent with
previous simulations which showed the possibility of disk formation
from major mergers (Springel, \& Hernquist, 2005; Robertson et al.
2006b).

\begin{figure*}
\includegraphics{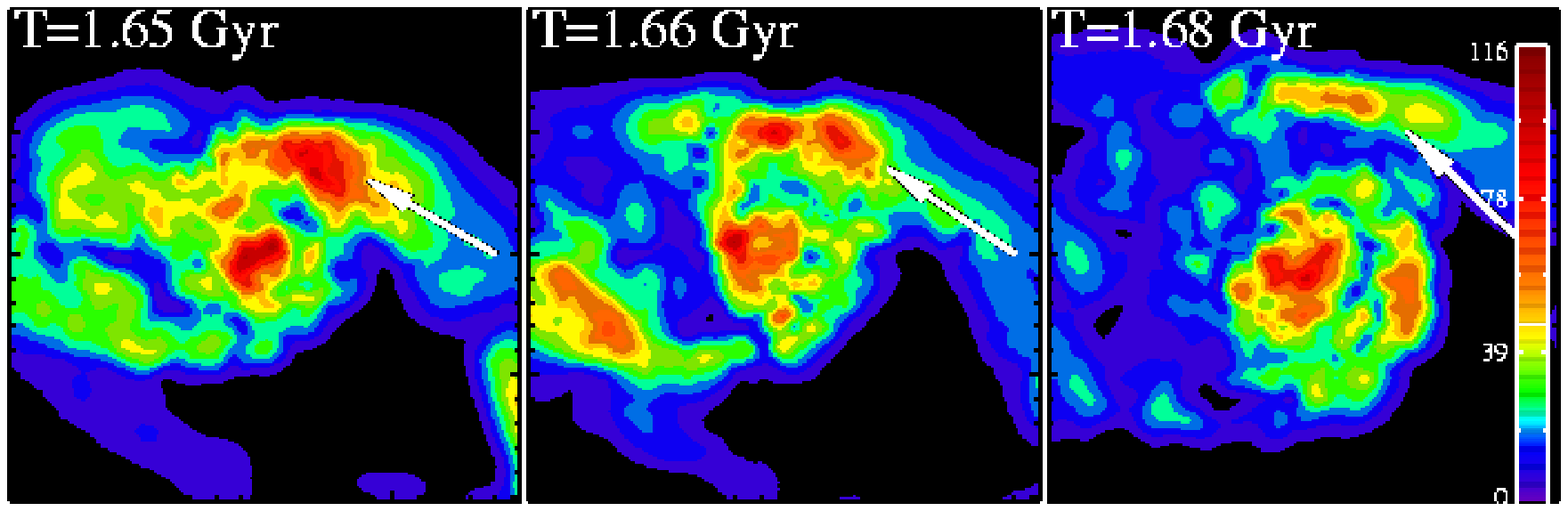}
\caption{Coplanar merger simulation, co-BH, soon after nuclear
coalescence. The molecular outflows occur preferentially in the plane
of the merger in a shell-like structure. An emergent outflow is
visible near the top of the nuclear region, as well as the remnant of
an older outflow in the bottom right of the first two panels. For
reference, the point of coalescence for model co-BH is T=1.5 Gyr.
\label{figure:vc3vc3h_2c_outflow}}
\end{figure*}

\begin{figure}
\includegraphics[angle=90,scale=0.35]{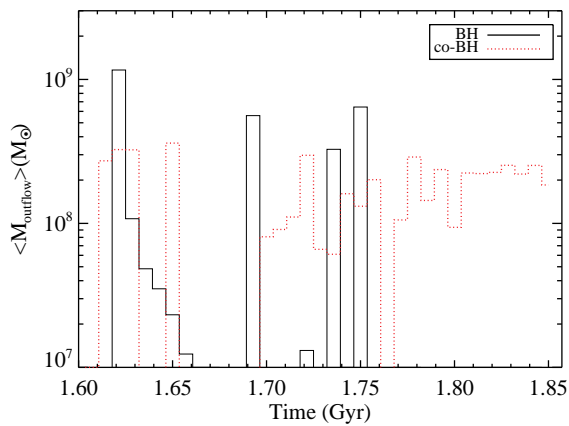}
\caption{Masses of detectable outflows as a function of time for model
BH and co-BH (thus varying only merger orbit angle) averaged over
three orthogonal sightlines The masses of a typical outflow are
comparable between models, though the coplanar model drives outflows
for longer periods of time.
\label{figure:imageable_clumps_mass2}}
\end{figure}

In Figure~\ref{figure:vc3vc3h_2c_outflow}, we show a typical outflow
from the coplanar model. Similar to the other models in
Table~\ref{table:ICs}, the primary molecular outflows occur at or near
the point of nuclear coalescence.  Unlike models sb/sbBH/BH, which
showed no preferred direction for outflows, the outflows in coplanar
model co-BH are seen to occur almost exclusively in the plane of the
merger with an expanding shell-like structure.\footnote{We caution
that the outflows seen here are unrelated to the ``shells'' and
``ripples'' seen around many merger remnants and ordinary ellipticals.
These features are stellar-dynamical in origin and are caused by
``phase-wrapping'' of stars (e.g. Quinn 1984; Hernquist \& Quinn 1987)
as tidal debris falls into the potential well of the remnant
(Hernquist \& Spergel 1992).  Indeed, gas associated with these shells
is on self-intersecting trajectories and will collect at the center of
the remnant (e.g. Weil \& Hernquist 1992, 1993).}  The fraction of
time in our simulations outflows are visible is dramatically increased
as well in this merger model. In the sb/sbBH/BH models, the outflows
typically involve loosely bound gas entrained in AGN winds which fall
back into the nuclear potential, tending the emission toward a
centrally concentrated morphology. Conversely, the outflows in the
coplanar model appear to continue expanding within a 6 kpc (radius)
region, resulting in an increasing half light radius throughout the
evolution of the post-merger galaxy. The continued presence of
detectable outflows in the coplanar model is seen explicitly in
Figure~\ref{figure:imageable_clumps_mass2}, where we show the mass of
detectable outflows for models BH and co-BH (thus varying only the
orientation angle) as a function of time.

The outflows are most dramatic for face on viewing angles as the
outflows are preferentially within the plane of the merger. While at
some inclined angles outflows continue to be visible, broadly, the
more edge-on the viewing angle becomes, the more difficult outflows
become to detect against nuclear emission. That said, outflows may be
preferentially detectable along these sightlines via spectral line
signatures which we will discuss further in
\S~\ref{section:lineprofiles}. Within the plane of the merger,
outflows can be viewed in outflow, or falling back into the nucleus
throughout the entire $\sim$200 Myr peak of the active phase
investigated here.

The peak black hole accretion rate and integrated energy deposited
into the ISM are similar for both the coplanar model and model
BH. Consequently, the typical physical conditions of the AGN
feedback-driven outflows are roughly the same in the two
simulations. In the coplanar model discussed in this section, the
typical mass of an outflow is a few $\times$10$^8$\msunend, similar to
the outflows observed in model BH which had a more generic inclination
angle for the progenitor galaxies.

\subsection{CO Half-Light Radius}
\label{section:halflight}

\begin{figure}
\includegraphics[angle=90,scale=0.35]{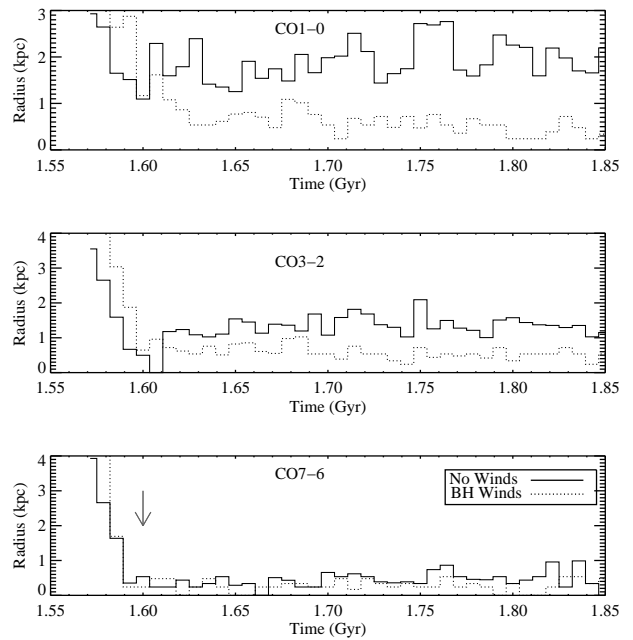}
\caption{CO halflight radius as a function of time for models with AGN
winds only (model BH) and with no winds (model no-winds) for
observable CO transitions CO (J=1-0), CO (J=3-2) and CO (J=7-6). The
emission from model BH with winds is uniformly compact in all CO
transitions, typically confined to a region of a few hundred pc. The
emission from model no-winds, without winds, is stratified, with a
characteristic CO (J=1-0) scale of $\sim$1.5-2 kpc, and becoming more
compact at higher lying transitions. The arrow in the third panel
denotes the point of nuclear coalescence. (See text for
details.)\label{figure:halflight_2models}}
\end{figure}

\begin{figure}
\includegraphics[scale=0.35]{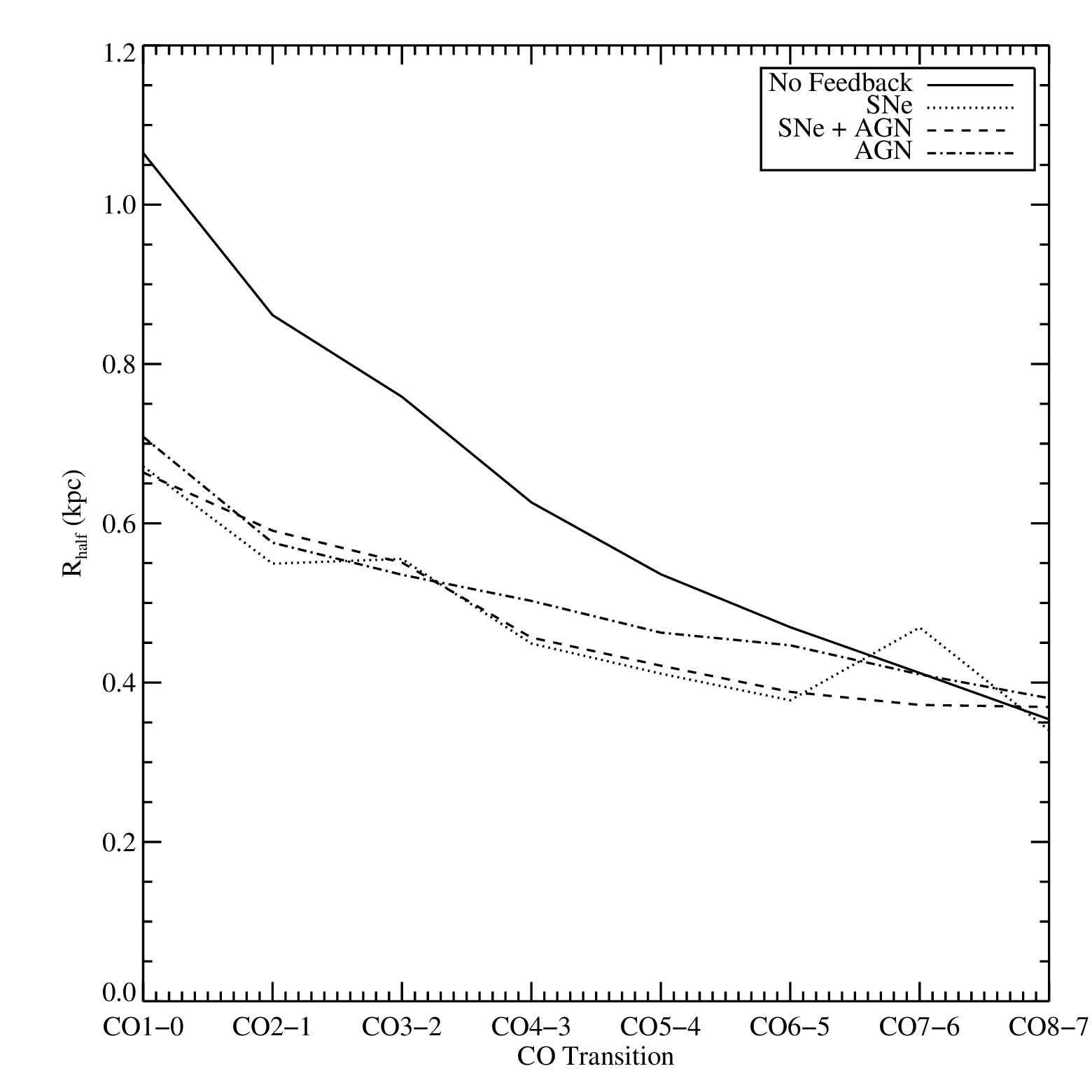}
\caption{Average halflight radius as a function of transition for
models BH, no-winds, sb and sbBH (thus varying only wind
parameters). The average halflight radius is calculated after nuclear
coalescence. The emission from the model without winds is relatively
stratified, with emission becoming more compact at higher
transitions. The models with winds also exhibit stratification, though
the halflight radius for the lowest transitions is about a factor of
two smaller than the models without winds.
\label{figure:halflight}}
\end{figure}

Galactic winds can play a role in the spatial extent of CO emission in
galaxy mergers.  After nuclear coalescence, the CO (J=1-0) emission in
the simulation without winds (model no-winds) is seen to be more
spatially extended than the emission from mergers with winds by a
factor of $\sim$2. The CO (J=1-0) emission from model no-winds has a
typical halflight radius of 1-2 kpc
(Figure~\ref{figure:halflight_2models}, top panel).  The CO emission
remains on these extended scales throughout the evolution of the
merger, and star formation simply consumes the available molecular gas
as fuel.

The predicted lack of extended CO (J=1-0) emission in galaxy mergers
with winds is consistent with observations. In a high resolution CO
survey of five mergers, Bryant \& Scoville (1999) found the primary CO
emission to be compact and concentrated within a 300-400 pc radius of
the nucleus in three sources, and within 1.6 kpc in all of the objects
in the sample. Moreover, high resolution observations and radiative
transfer modeling of local mergers have confirmed the compact nature
of CO emission in mergers (Downes \& Solomon, 1998). Winds associated
with growing black holes and/or starbursts may help to explain this.

In the models with winds, as the galaxies merge, the sudden inflow of
gas into the nucleus rapidly drives up the nuclear star formation
rate, as well as the accretion rate of the black hole(s). The feedback
associated with star formation and the central AGN is effective at
heating the gas in the nuclear regions, and dispersing lower density
CO (J=1-0) gas, leaving behind only a dense core of cool gas in the
central regions in the winds models.  The average CO (J=1-0) half
light radius after the nuclei have coalesced for the models with SNe
and/or AGN is $\sim$700 pc. This was also seen in Figure
~\ref{figure:sb10BH_morph} where we showed that the major CO (J=1-0)
intensity for the model with SNe and AGN winds originates in a compact
region, in agreement with observations of local ULIRGs (Bryant \&
Scoville, 1999; Downes \& Solomon, 1998).

In principle, these model results can be directly tested through
observations of higher lying CO lines from local galaxies. The middle
and bottom panels of Figure~\ref{figure:halflight_2models} show the
halflight radius as a function of time for models BH and no-winds for
the CO (J=3-2) and CO (J=7-6) transitions. These transitions probe
denser gas than CO (J=1-0), and have a critical density of
$\ga$10$^{4}$-10$^5$\cmthree. Figure~\ref{figure:halflight_2models}
shows that the emission from the model with no winds is stratified
such that tracers of dense gas [e.g. CO (J=7-6)] are more compact by
nearly an order of magnitude than tracers of lower density gas
[e.g. CO (J=1-0)]. In Figure~\ref{figure:halflight} we show the mean
half light radius for each model plotted as a function of CO
transition. Here, we have averaged the half light radius after nuclear
coalescence in each simulation. The CO emission in the model without
winds becomes more compact in higher lying transitions (e.g. those
with higher critical densities).  In the models with SNe and/or AGN
winds, the CO emission shows a similar behavior, but the halflight
radius for the lowest transitions is about a factor of two smaller
than in the models without winds.

With these predictions, we caution that specific values quoted here
are dependent on a number of assumptions included in parameter choices
(such as, for example, the specific multi-phase ISM breakdown
[e.g. Springel \& Hernquist, 2003] or constant CO abundances), and
thus the normalization of the plots presented are to be taken within
those confines. However, the general trends are likely to be more
robust. In this sense, the relatively shallow slope in half-light
radius versus transition (Figure~\ref{figure:halflight}) remains a
valid observational test for the role of galactic winds in shaping
observed CO spatial extents. Finally, it is important to note that the
weak trend in average halflight radius as a function of transition in
Figure~\ref{figure:halflight} is an ensemble-averaged trend over three
orthogonal sightlines and numerous points in the post-coalescence
evolution of the galaxies. Thus, observations examining this
relationship must select samples of galaxies after the progenitors
have coalesced - in other words, after the merger has only a single
nuclear emission peak in the CO morphology.

\section{Signatures of Winds in CO Line Profiles}
\label{section:lineprofiles}

\subsection{Overview}
\label{section:lineoverview}

\begin{figure}
\scalebox{1.2}{\plotone{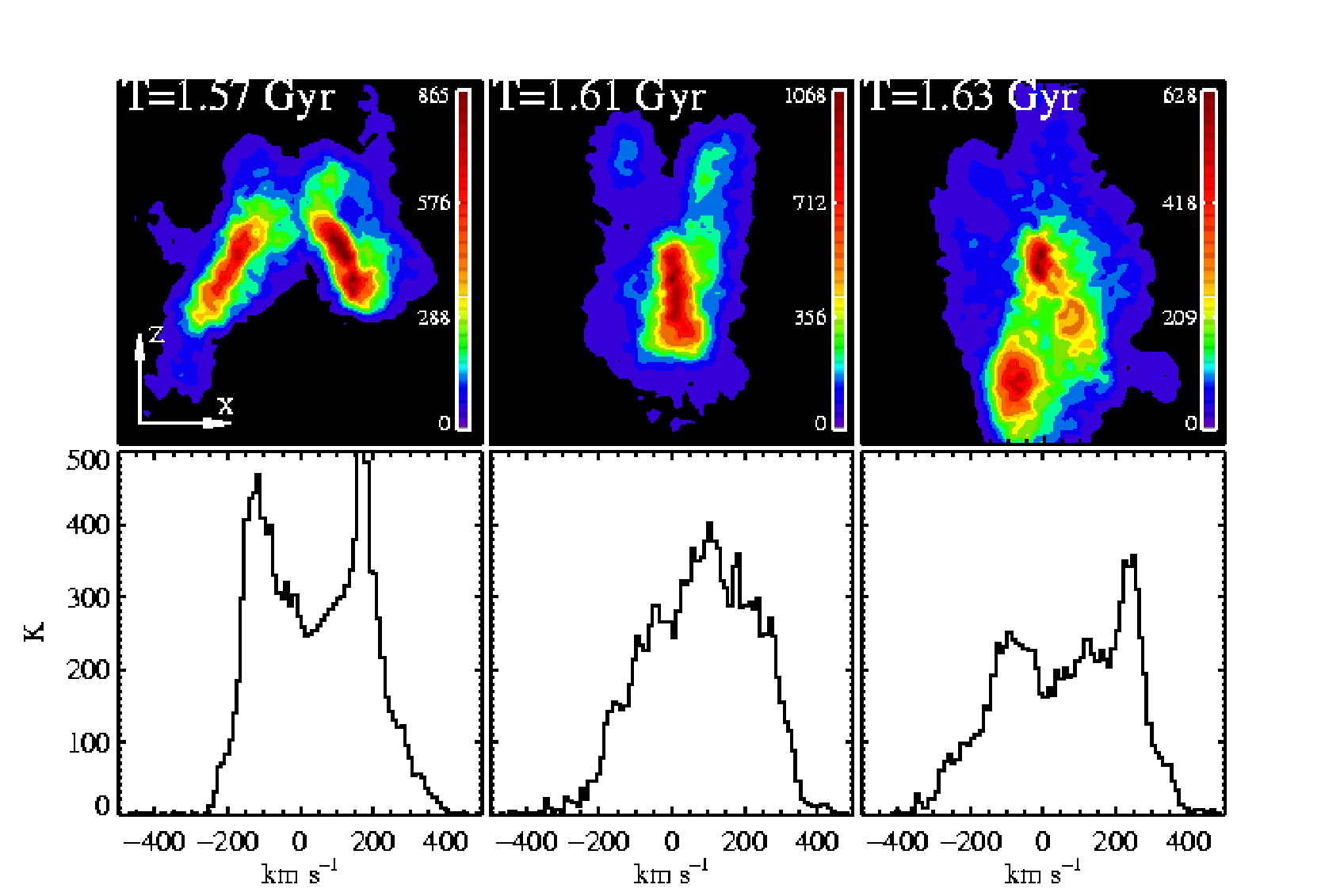}}
\caption{{\it top:} CO (J=1-0) emission contours of three snapshots
during the merger simulation with winds from black holes (model
BH). {\it bottom:} CO (J=1-0) spectra of snapshots, viewing the
emission from the -$\hat{z}$ direction (coordinate axes in top left
panel). The CO emission lines early in the merger are
characteristically double-peaked, where each peak corresponds to
emission from the individual nuclear regions of the progenitor
galaxies (first column). As the merger advances, the emission lines
are typically well-described by a single Gaussian (middle
column). There are excursions from this trend, however, as outflowing
gas can give rise to a secondary peak in the emission line red or
blueshifted from the systemic velocity of the galaxy (third column;
this outflow and line profile was also presented in Narayanan et
al. 2006a). The emission contours each have their own scale on the
right, to facilitate color contrasts. For reference, the time of
nuclear coalescence is T=1.6 Gyr. The units of emission are in
velocity-integrated intensity (\kkms). The panels are 12 kpc on a
side.
\label{figure:map_spectra}}
\vspace{.15cm}
\end{figure}

\begin{figure}
\scalebox{0.9}{\rotatebox{90}{\plotone{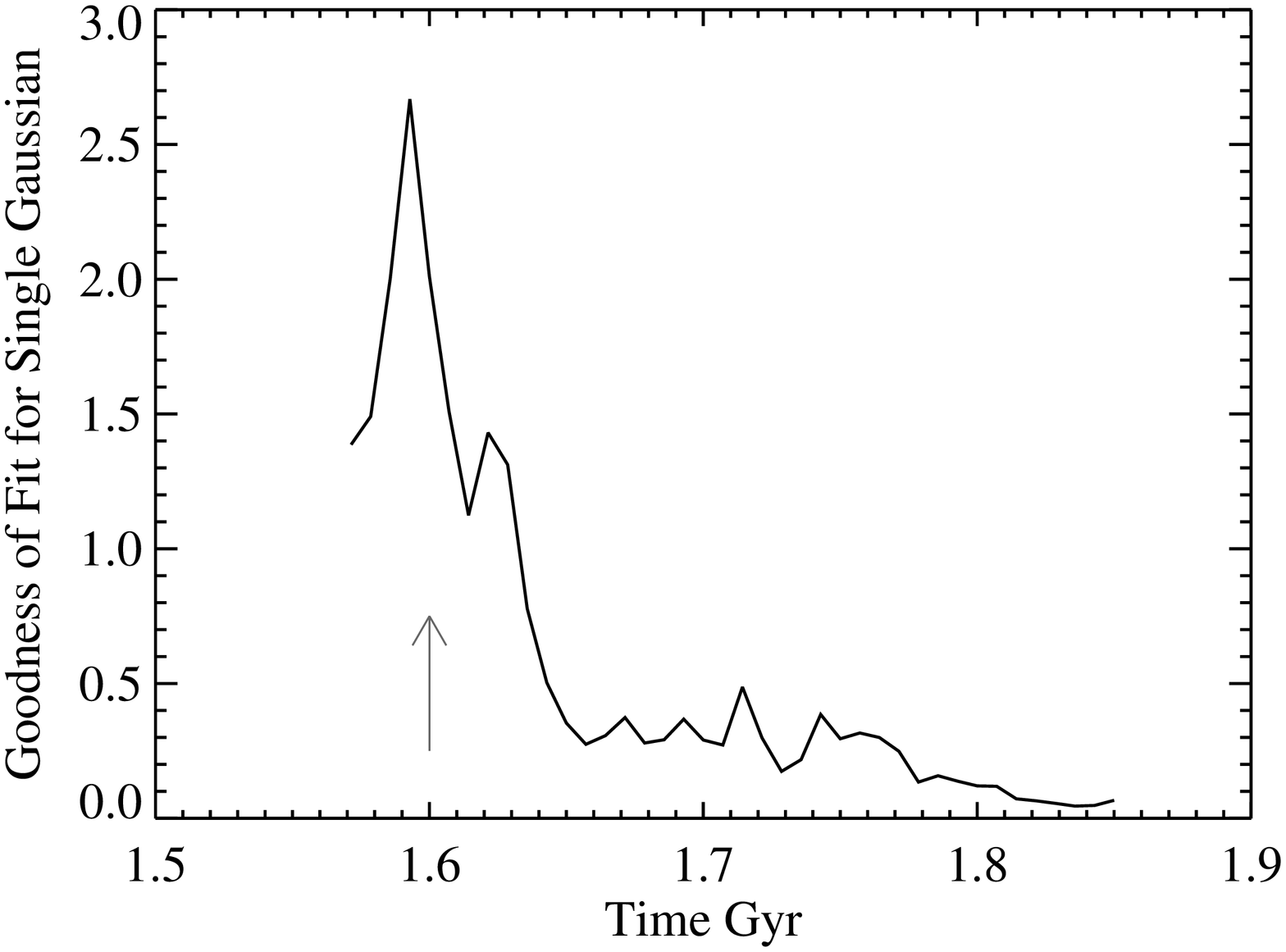}}}
\caption{Goodness of fit for a single Gaussian fit to spectra from
model BH as a function of time. When the merger is prior to and near
coalescence, and the gas is highly dynamical, and a single Gaussian
provide a relatively poor description of the emission line profile. As
the gas virializes after coalescence, a single Gaussian better
represents the emission line profile. The goodness of fit is taken
here as the median of the square of the residuals. The arrow denotes
the time of nuclear coalescence.
\label{figure:chi}}
\end{figure}

Generically, when the nuclei of the progenitor galaxies are still
resolvable, prior to coalescence, the unresolved CO emission line
exhibits multiple emission peaks corresponding to the nuclear regions
of each galaxy (e.g. Greve \& Sommer-Larsen, 2006). When the molecular
gas in galaxy mergers becomes dynamically relaxed, the CO emission
line is reasonably well represented by a single broad
Gaussian. Alongside this general trend, at various times throughout
the evolution of the merger, secondary (and potentially, though less
commonly, tertiary and quaternary) peaks can be seen superposed on the
broad CO Gaussian-like line. In Figure~\ref{figure:map_spectra}, we
show a representative evolution of CO spectral lines, and how they
correspond to the CO morphology. These secondary peaks superposed on
the broad Gaussian originate in massive CO gas clumps moving along the
line of sight. In Figure~\ref{figure:map_spectra}, the secondary peak
in the third panel represents emission from the outflowing clump
visible in the corresponding CO contour image.

While these superposed secondary emission peaks are a general feature
of all models in our simulations, the physical origin of these peaks
is dependent on the strength of the winds included in a given merger
model. In particular, secondary emission peaks with velocities greater
than the circular velocity typically represent loosely bound gas which
has not virialized (peaks at velocities lower than the circular
velocity may also be loosely bound gas, though they may also represent
gas in rotational support; for brevity we hereafter refer to secondary
peaks at velocities greater than the circular velocity as 'high
velocity peaks'). As we will show more quantitatively, in models which
include winds, this high velocity gas is typically entrained in
outflow during periods of activity (e.g. elevated star formation rate
or black hole growth), and thus give rise to high velocity peaks in
the emission line representative of outflowing gas. This is analogous
to the situation seen in the CO morphologies presented in
Figure~\ref{figure:outflows_differentmodels} when loosely bound gas
simply fell into the central potential in the model without winds, but
was entrained in outflow in models with winds (here, models sb, sbBH
and BH).  This effect of winds on CO emission profiles was first shown
explicitly in similar calculations by Narayanan et al. (2006a).  This
work demonstrated that once the galaxies merged, large columns of gas
outflowing along the line of sight can leave their imprint on the
broad single-Gaussian emission line characteristic of galactic CO
emission by introducing an additional superposed emission peak, red or
blueshifted at the line of sight velocity of the outflow.

The relative significance of high velocity peaks compared to the main
emission line decreases as the gas virializes in the merger. To
illustrate this, in Figure~\ref{figure:chi} we plot the evolution of
the goodness of fit (taken here as the median of the square of the
residuals) for a single Gaussian fit to the emission lines from model
BH, although the results are common across all models. The impact of
winds in driving high velocity peaks decreases as the merger
progresses.

In our models, high velocity peaks in CO emission lines are typically
representative of clumps of gas moving along the line of sight with
masses $\ga$5$\times$10$^{7}$\msunend.  Gas clumps with masses much
less than this are typically not able to emit strongly enough to be
seen in the emission spectrum over the main Gaussian.  In the
remainder of \S~\ref{section:lineprofiles}, when we discuss clumps of
outflowing and infalling gas, we will be referring to those above this
minimum mass that are able to impact the emission line profile.

Because high velocity peaks appear in our simulations even in
spatially unresolved observations, they may be a powerful diagnostic
for AGN and starburst induced outflows both in the local Universe and
at earlier epochs. However, at this point in the discussion, there are
open questions regarding the uniqueness of this line profile, as well
as how these secondary peaks specifically relate to starburst and AGN
activity. In the remainder of this section, we attempt to quantify the
use of molecular emission line profiles as a diagnostic for outflowing
gas in galaxy mergers.

\subsection{The Origin of High Velocity Peaks in CO Emission Profiles}

Emission lines representing gas with a strong line of sight velocity
component can originate from either outflowing or infalling gas. As
demonstrated by Narayanan et al.  (2006a), AGN feedback-driven
molecular outflows can generate high velocity secondary peaks in the
CO emission lines when the outflow has a significant line of sight
velocity component. Any gas falling into the central potential with a
negative radial velocity with respect to the center of mass of the
system (e.g gas that is in infall, or pre-coalescence mergers) may
also drive high velocity peaks in the CO emission lines, and are thus
degenerate with high velocity peaks that originate in outflows.  It is
therefore necessary to establish the uniqueness of high velocity peaks
in CO emission lines with respect to an actual outflow or infall
origin. Any attempt to break this degeneracy requires a more detailed
knowledge of the physical properties and observational characteristics
of the galaxy during periods when high velocity peaks are observable.

In order to assess how the evolutionary state of the galaxy merger may
determine the origin of high velocity peaks (i.e. whether they owe to
outflowing or infalling gas), we investigate the physical conditions
during times when the galaxy exhibits high velocity peaks.  To this
end, we created synthetic CO (J=1-0) spectra along three orthogonal
sightlines for the models in Table~\ref{table:ICs}. These spectra were
made by integrating over the entire CO emitting area.  To identify
candidate emission lines with high velocity peaks driven by winds (as
opposed to rotation), we require the emission line to have a peak near
the systemic velocity of the galaxy, and an additional peak with
velocity greater than $\pm$160 \kmsend.  This fiducial requirement is
set in place to filter out multiple emission peaks that arise from
rotation as $\sim$160 \kms is the circular velocity of the post-merger
galaxy at the spatial extent of the \htwo \ gas, and thus the maximum
velocity of rotating gas. 
 We consider all high velocity peaks whose peak intensity is at least
20\% the peak line intensity.

Upon identifying a line with a high velocity peak, we then determine
whether the clump of gas generating the high velocity peak is
outflowing or infalling. In order to simplify the analysis, in the few
cases with multiple high velocity peaks, we focused on the peak with
highest velocity. In Figure~\ref{figure:vc3vc3e_2b_outflowinfall}, we
plot a histogram of the number of high velocity peaks seen as a
function of time in the simulation with black hole winds only (model
BH), summed over 3 orthogonal lines of sight. We have further broken
the line profiles down into having an outflow or infall origin.  As a
reference for the evolutionary state of the merger, we have
overplotted the SFR, black hole accretion rate, and bolometric
luminosity.

Figure~\ref{figure:vc3vc3e_2b_outflowinfall} demonstrates that while
the number of high velocity peaks is roughly constant after the major
merger, the origin of these line profiles is a function of the
evolutionary state of the galaxy. During phases of heavy black hole
growth, the majority of the high velocity peaks owe to outflowing gas.
Towards the end of the quasar phase, when much of the blowout has
already occurred, gas raining from outer parts of the galaxy give rise
to multi-peaked profiles with an infall origin. Conversely, similar
analysis from the model without winds shows a more random origin for
high velocity peaks in the emission profiles, reflecting the more
randomly oriented velocity vectors of loosely bound clumps. This is
explicitly shown in Figure~\ref{figure:vc3vc3e_noc_outflowinfall},
where we show the origin of high velocity peaks in a merger without
winds compared with the bolometric and IR luminosity.

Another way of saying this is that the winds in model BH do not {\it
cause} the high velocity peaks to exist in the CO spectra. Rather
winds typically entrain the clumps of gas behind the high velocity
peaks in outflow, thus giving physical significance to secondary
peaks. This is not the case in the model with no winds, where the
molecular gas representative of the high velocity peaks has random
radial velocities with no preferred direction
(Figure~\ref{figure:vc3vc3e_noc_outflowinfall}). {\it High velocity
peaks are seen in models both with and without winds; in wind models,
though, these peaks preferentially owe to outflowing gas during peaks
of starburst activity and black hole growth. }

Observationally, many of the criteria that are used in creating
samples of local galaxy mergers may select galaxies which are at the
peak of their starburst or AGN activity. High velocity peaks in the
emission profiles (e.g. higher than the circular velocity of the host)
observed in these galaxies may be likely, then, to owe to outflowing
clumps of gas (Figure~\ref{figure:vc3vc3e_2b_outflowinfall}). For
example, the extreme bolometric and IR luminosities trace not only the
prodigious star formation owing to the merger induced starburst, but
also reflect a large contribution from the growing supermassive black
hole. The winds associated with this black hole growth are behind much
of the outflowing molecular gas during the most luminous phase of the
galaxy's evolution. Similarly, IR colors which select objects with
embedded AGN, such as the 25$\mu$m/60$\mu$m flux ratio, may
preferentially show objects which owe their outflow profile to
outflows as opposed to infalling gas clumps. Finally, the hard X-ray
flux is dominated by the central AGN, and thus traces the general
growth of the central black hole. Our models suggest that the phase
during which the X-ray luminosity is at its greatest will be marked by
high velocity peaks in the CO emission originating in outflowing
molecular gas.

\subsection{Distinguishing Characteristics}
\label{section:sbagnlines}

\begin{figure*}
\scalebox{1.}{\rotatebox{90}{\plotone{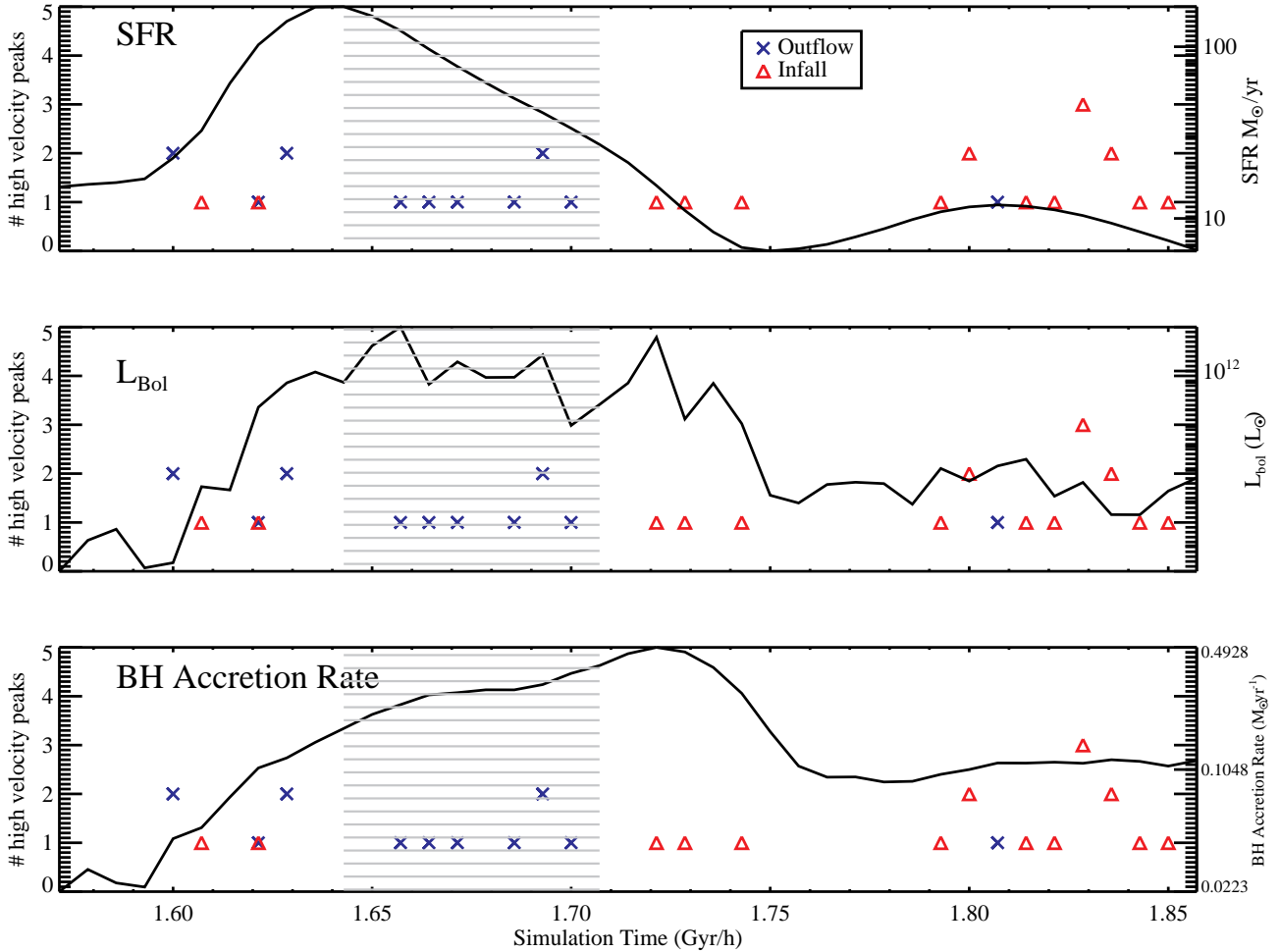}}}
\vspace{1cm}
\caption{Physical and observable properties of model BH (model with BH
winds) as a function of simulation time. Additionally plotted are a
histogram of the number of secondary high velocity peaks seen, and the
physical origin of the peak. The blue crosses are number of high
velocity peaks associated with an outflow (summed over 3 orthogonal
viewing angles), and the red triangles are number of high velocity
peaks originating from infalling gas. Infalling gas is seen both early
and late in the merger's evolution. During periods of elevated
activity (e.g. increased SFR or black hole accretion), most of the
high velocity peaks owe to outflowing gas (hatched region). The left
axis is associated with the histogram of high velocity peaks, and the
right axis with the physical or observable quantity overplotted in
that panel. For reference, the time of coalescence is T=1.6
Gyr. \label{figure:vc3vc3e_2b_outflowinfall}}
\vspace{.1cm}
\end{figure*}

\begin{figure*}
\scalebox{1.}{\rotatebox{90}{\plotone{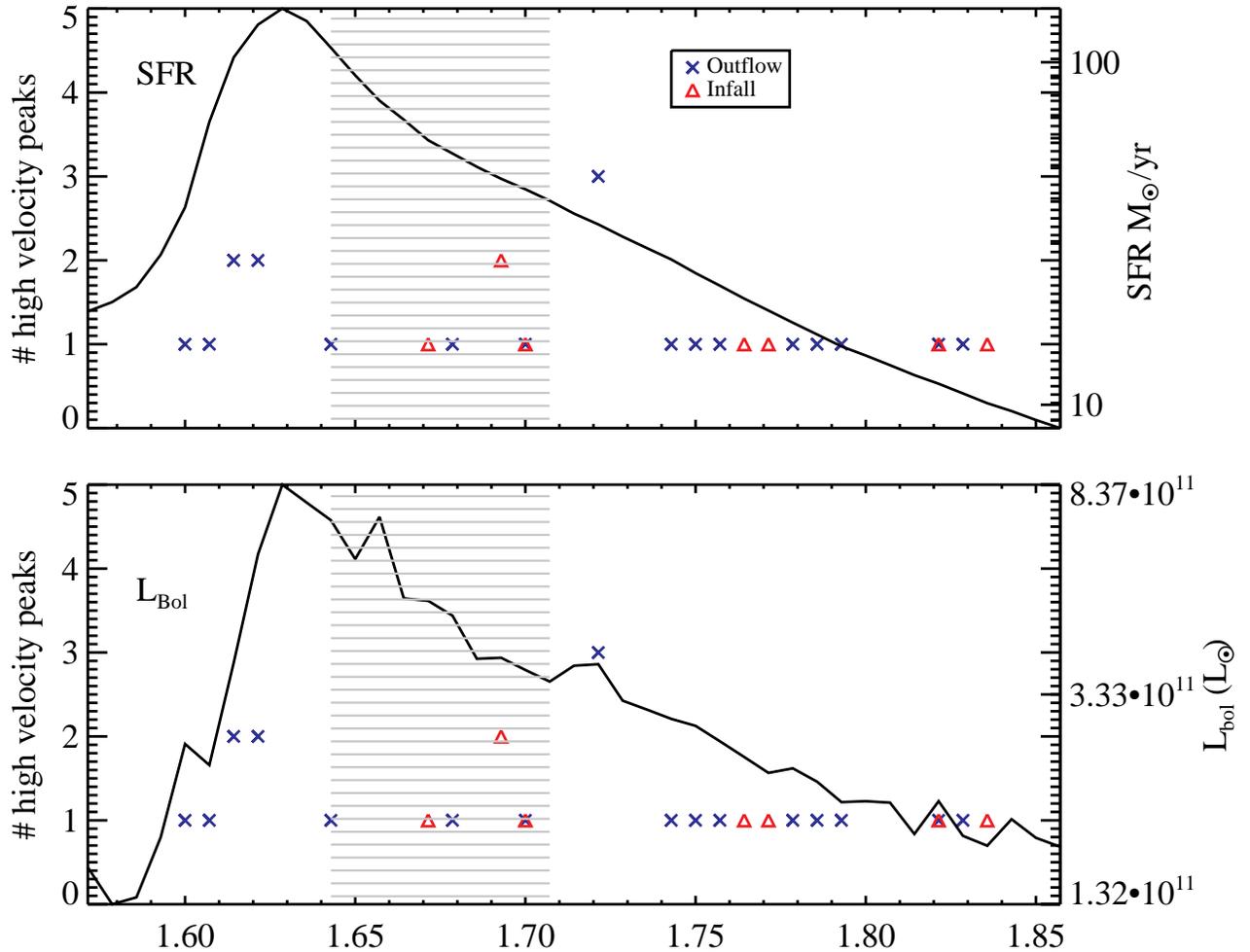}}}
\vspace{1cm}
\caption{ SFR and bolometric luminosity from model no-winds as a
function of simulation time. Additionally plotted are a histogram of
the number of secondary high velocity peaks seen, and the physical
origin of the peak. The blue crosses are number of high velocity peaks
associated with an outflow (summed over 3 orthogonal viewing angles),
and the red triangles are number of high velocity peaks originating
from infalling gas. No apparent trends are seen reflecting the
randomly oriented radial velocity vectors driving the high velocity
peaks in the no-winds model. The hatched region covers the same time
period as that in Figure~\ref{figure:vc3vc3e_2b_outflowinfall} just as
a comparative reference with that plot.  The left axis is associated
with the histogram of high velocity peaks, and the right axis with the
physical or observable quantity overplotted in that panel. For
reference, the time of coalescence is T=1.6
Gyr. \label{figure:vc3vc3e_noc_outflowinfall}}
\vspace{.1cm}
\end{figure*}

A natural question is whether or not one can observationally infer the
nature of high velocity peaks in CO emission lines as discussed in
\S~\ref{section:lineoverview}. Specifically, high velocity peaks can
originate in both wind-driven outflows, as well as owing to dynamical
effects in merging galaxies
(e.g. Figures~\ref{figure:vc3vc3e_2b_outflowinfall}
and~\ref{figure:vc3vc3e_noc_outflowinfall}).  Distinguishing
characteristics between the various origins for high velocity peaks
may be evident when considering limiting velocities for the high
velocity peak much greater than the virial
velocity. Figures~\ref{figure:vc3vc3e_2b_outflowinfall}
and~\ref{figure:vc3vc3e_noc_outflowinfall} showed the presence of all
high velocity peaks beyond a limiting velocity being set at the virial
velocity of the galaxy (here, $\sim 160$ \kmsend). When considering
the origin of high velocity peaks driven by dynamical effects, few
emitting clumps of gas are able to regularly sustain velocities much
greater than the escape velocity of the galaxy (or $\sqrt{2} \times
V_c$). In contrast, when included, winds are able to drive outflows at
higher speeds.

\begin{figure}
\scalebox{1}{\rotatebox{90}{\plotone{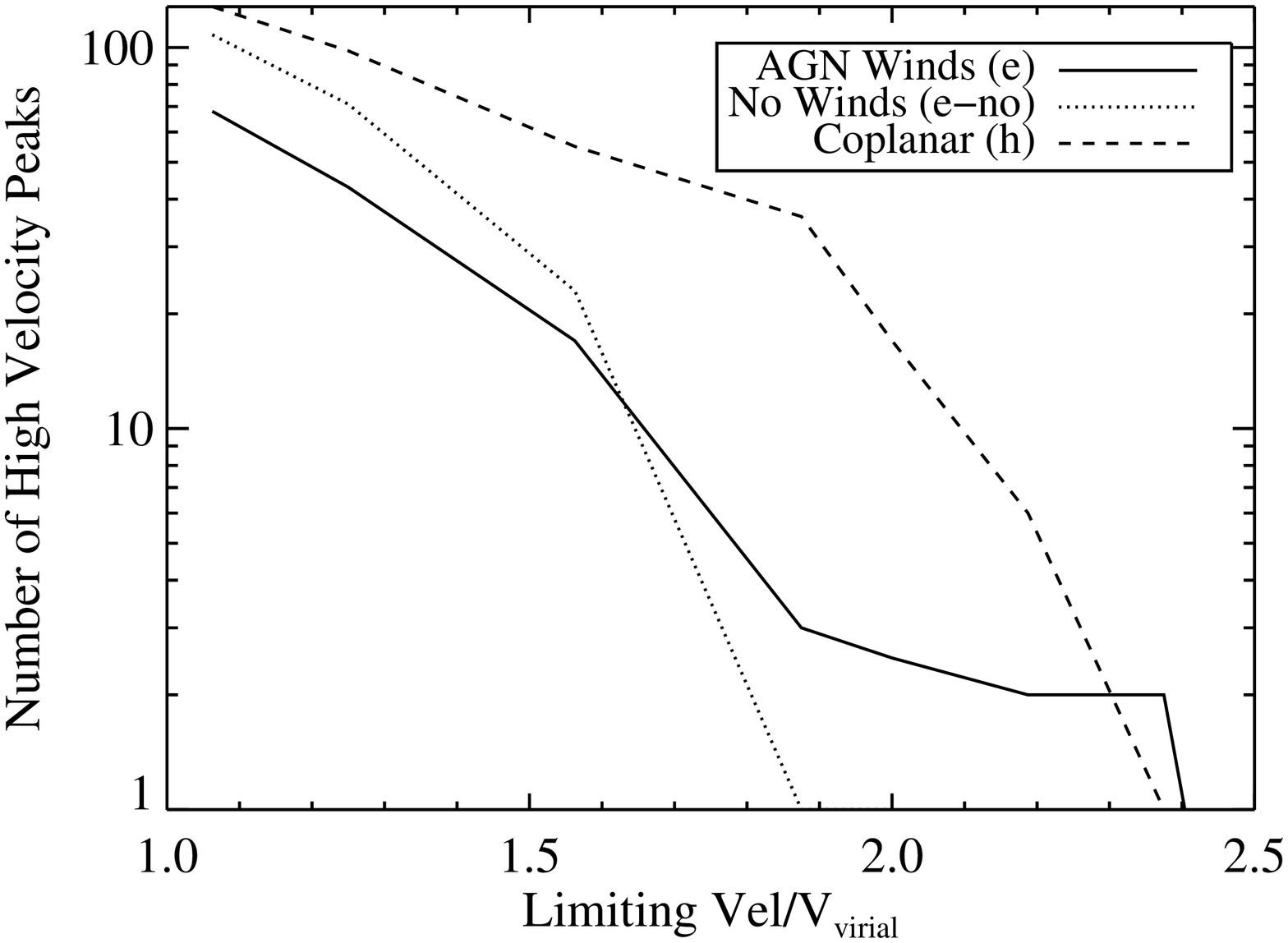}}}
\caption{Number of high velocity peaks as a function of minimum
velocity for the high velocity peak summed over three orthogonal
sightlines. Solid line is model with AGN winds (model BH) and dashed
line is model with no winds. High velocity peaks in model without
winds are driven by random dynamics of clumps of gas, and thus are
unable to sustain velocities much greater than the virial velocity of
the host galaxy. In contrast, clumps of gas entrained in AGN winds are
seen at higher velocities, giving rise to more high velocity peaks at
the highest velocities. One may be able to discriminate between high
velocity peaks with an outflow origin versus random kinematics when
restricting the limiting velocity $\ga$1.7 times the virial
velocity.\label{figure:vspike_noutflows}}
\end{figure}

We show this effect quantitatively in
Figure~\ref{figure:vspike_noutflows} where we plot the number of high
velocity peaks as a function of limiting velocity for both two models
with AGN winds (models BH and co-BH) and the model with no winds. At
velocities greater than $\sim$1.7 times the virial velocity, the
number of high velocity peaks driven by purely dynamical effects in
model no-winds drops rapidly while winds from the AGN in model BH and
co-BH are still able to drive clumps in outflow at these velocities.
The range of the numbers of high velocity peaks in the models with AGN
winds owes to a merger-orientation dependent range in instantaneous
black hole feedback energies. More efficient fueling of the central
AGN in the coplanar model allows for a stronger impact of the winds on
the molecular line profiles
(cf. \S~\ref{section:morphorientation}). In either case, both models
which consider AGN feedback show that AGN feedback-driven winds can
drive high velocity peaks at larger velocities than dynamical effects
can typically account for.  In this sense, one may be able to
discriminate between high velocity peaks with an outflow origin versus
random kinematics when restricting the limiting velocity of the peaks
to $\ga$1.7 times the virial velocity of the galaxy. It is important
to note, however, that this is specific to the single merger with no
winds examined in this work, and may not necessarily be general.

The dispersion seen between the no winds model and the models with AGN
winds (BH and co-BH) in Figure~\ref{figure:vspike_noutflows} is in
principle a lower limit.  The galaxies in
Figure~\ref{figure:vspike_noutflows} are relatively low mass -
$M_\star \sim10^{11}$\msunend. Cox et al. (2007) found that the energy
input by accreting black holes in galaxy mergers increases as a
function of galaxy mass for a variety of progenitor gas fractions
(ranging from 0.05 to 0.8) and initial merger angles. Specifically,
these authors found $E_{\rm BH}/f_g \propto M_\star^{1.13}$,
suggesting that higher mass galaxies have the capacity to drive CO
peaks at even higher speeds compared to their virial velocity. As an
example, a merger with similar gas fraction and orientation angle as
model BH, though with final stellar mass $\sim 10^{12}$ would have
roughly an order of magnitude greater black hole energy input (Cox et
al. 2007).

Figure~\ref{figure:vspike_noutflows} suggests that the effect of AGN
feedback-driven winds on CO emission line profiles should be
discernible by comparing a merger's virial velocity to observed high
velocity features in the CO line profile.  A comparably clear (and
observationally more tractable) test may be an analysis of the
velocity offset of high velocity peaks with respect to the centroid
velocity of the emission line. When examining high velocity peaks
defined with respect to the centroid velocity, we find the curves
presented in Figure~\ref{figure:vspike_noutflows} are robust. This
owes to a narrow distribution of centroid velocities in all the
models, with a typical standard deviation in the distribution $\sigma
\sim 20$\kmsend.

\section{Discussion}
\label{section:observations}

Many of the model results presented here have the potential to be
directly tested in the near future, either with existing datasets, or
with observations from millimeter-wave interferometers such as CARMA
and PdBI, and high resolution submillimeter telescopes such as the
Submillimeter Array (SMA) or Atacama Large Millimeter Array (ALMA).

First, it is worth briefly revisiting the evolutionary status of the
galaxies studied here, and placing them into a larger context. In a
merger-driven scenario for quasar formation, gaseous inflows fueled by
tidal torquing on the gas (Barnes \& Hernquist 1991, 1996; Hernquist,
1989) fuel nuclear starbursts (Mihos \& Hernquist, 1996), allowing the
galaxy to be viewed as a dusty starburst system similar to local
ULIRGs (Chakrabarti et al. 2007a). The gaseous inflows fuel central
black hole growth (Hopkins et al. 2005a-d; 2006a-d; 2007h; Di Matteo
et al.  2005; Springel et al. 2005a) which makes the galaxy first
visible as a ULIRG with warm infrared colors (Chakrabarti et
al. 2007a), and then as an optically selected quasar (Hopkins et
al. 2005a-d; 2006a-d; 2007h).  A key physical process in driving the
host galaxy along this sequence is galactic winds. Namely, winds
driven by AGN feedback (as well as nuclear starbursts) contribute to
clearing the veil of obscuring gas and dust surrounding the central
black hole, and allow the quasar to become visible across numerous
sightlines (Hopkins et al. 2005a). These winds are most prominent
during the $\sim$200 Myr associated with major black hole growth. The
models presented in this work therefore examine simulated galaxy
mergers during this time period of heightened wind activity, starting
near nuclear coalescence, and ending slightly after the quasar light
curve has declined. During this time, the galaxy is seen to have
increased IR, X-ray and $B$-band luminosities, indicative of the
starburst event and central black hole growth
(Figure~\ref{figure:vc3vc3e_2b_outflowinfall}). In this sense, these
simulations are designed to provide analysis for local galaxies
selected for nuclear activity via elevated X-ray and IR fluxes
(e.g. ULIRGs).

Second, we have shown that the effects of winds can shape the
molecular morphology of galaxy mergers. For example, distinct
differences are seen in the simulations with winds versus without when
considering the spatial extent of the CO emitting gas
(\S~\ref{section:halflight}). The predicted weak evolution of average
half-light radius as a function of transition for evolved mergers can
be tested in the CO (J=1-0), (J=2-1), (J=3-2) and (J=7-6) transitions
with current technology using a millimeter-wave interferometer
(e.g. CARMA, PdBI) and the SMA.

Perhaps the most dramatic effects of galactic winds on the CO
morphology lie in signatures of entrained molecular outflows in
emission contour maps. CO outflows driven by starburst-dominated winds
have been imaged in a handful of nearby galaxies. Recent mapping of
starburst galaxies M82 and NGC 3256 have shown that molecular gas can
indeed survive in outflows entrained in starburst-driven winds (Walter
et al, 2002; Sakamoto et al. 2006). Both galaxies are known to be
interacting systems, the latter a major merger. The outflow masses are
recorded to be $\sim$10$^7$-10$^8$\msunend, consistent with the model
results presented in \S~\ref{section:morphology}. The outflows have
been detected in CO (J=2-1) emission (peaking at CO J$\sim$3 or 4;
Wei\ss \ et al. 2005b). The outflow observed in M82 appears to be
distributed in a roughly spherical halo with radius $\sim$1.5
kpc. This outflow may correspond with an H$\alpha$ outflow, as well as
M82's dust halo (Alton, Davies \& Bianchi, 1999; Walter et
al. 2002). This type of morphology is seen along some sightlines in
the models BH and co-BH, alongside the more collimated outflows
characteristic of e.g.  Figure~\ref{figure:outflows_differentmodels}.

Outflows which may be driven by a central AGN have been imaged in
local galaxies as well. Recent SMA observations of NGC 6240 by Iono et
al. (2007) have revealed an outflow in CO (J=3-2) morphologically
similar to the outflow seen in
Figure~\ref{figure:vc3vc3e_outflow}. NGC 6240 is a major merger near
final coalescence with two central X-ray sources, which {\it Chandra}
observations have suggested are a binary AGN (Komossa et al, 2003). In
this sense, the outflow imaged in NGC 6240 occurs when our models
suggest the most massive outflows may be imageable
(\S~\ref{section:relativestaragnmorph}). The mass of the outflow in
this system is estimated to be $\sim$5$\times$10$^8$\msunend,
consistent with the typical masses entrained in AGN winds in our
simulations. Similar morphological features to those associated with
molecular outflows have been detected in local LIRG/ULIRG NGC 985,
though the interpretation of these features is still uncertain
(Appleton et al. 2002).

While the number of outflows directly imaged in CO remains small and
limited to only relatively recent observations, outflows traced in
absorption (e.g. Na I D lines) have a much more extensive detection
history and associated literature. Absorption line outflows have been
detected in the UV and optical in both nearby starburst galaxies/AGN
(Heckman et al. 2000; Martin, 2005; Rupke \& Veilleux, 2005; Rupke,
Veilleux \& Sanders, 2005a-c), as well as high-redshift galaxies
(Pettini et al. 2002; Shapley et al. 2003) and quasars (Ganguly et
al. 2006; Hamann, Barlow \& Junkkarinen, 1997; Misawa et al. 2005,
2007a,b; Narayanan et al. 2004; Trump et al. 2006; Wise et al.
2004). HI has also been seen in outflow in active galaxies via
absorption lines (Morganti, Tadhunter \& Oosterloo; 2005). In addition
to the neutral absorption line outflows detected, many outflows are
relatively high ionization lines (e.g. CIV, NV; Hamann et al. 1997;
Narayanan et al. 2004). In this sense, it appears as though these may
be different populations of outflows from the cold, neutral emission
line outflows discussed in this work. Interestingly, in the few
systems with CO outflow candidates (as evidenced by their morphology),
there is an apparent coincidence between the velocity offsets in the
molecular outflows and absorption line outflows. For example, the
starburst-driven outflow observed in local merger NGC 3256 by Sakamoto
et al. (2006) has a velocity offset (from the systemic, which in the
case of CO, typically corresponds to the nuclear emission) of
$\sim$300 km/s, while Heckman et al. (2000) showed a blueshifted Na I
D absorption line of velocity $\sim$300 km/s. Similarly, Appleton et
al. (2002) found two Ly$\alpha$ and NV absorbers in NGC 985 at
coincident velocities as those which CO gas is seen projected against
the nucleus. The data bring up the interesting speculative suggestion
that perhaps the absorption line outflows seen in these systems are
physically associated with the CO emitting outflows. If so, it may be
that the neutral and ionized outflows exist more toward the outer
envelopes of the entrained complexes of molecular clouds as the
conditions toward the central regions of these flows are extremely
cold and dense ($\sim$10-30 K, $\sim$10$^5$ \cmthree;
Figure~\ref{figure:vc3vc3e_outflow}). A full investigation into the
potential relationship between UV and optical absorbers and CO
outflows is beyond the scope of this work, but studies into possible
physical motivations for absorption line outflows within the context
of these models are underway.

In principle, unresolved emission line profiles from galaxies are more
easily obtainable in large numbers than the high-resolution maps
necessary to directly image outflows. Brief inspection by eye suggests
that at least some local galaxies appear to espouse the high velocity
peaks discussed in \S~\ref{section:lineprofiles} (e.g. Narayanan et
al. 2005; Sanders Scoville \& Soifer, 1991; Solomon, Downes \&
Radford, 1992; Yao et al. 2003). As previously discussed, robustly
determining the physical mechanism driving high velocity peaks is
difficult, although tests similar to
Figure~\ref{figure:vspike_noutflows} may be possible with some prior
knowledge of the host circular velocity.  It is interesting to note
that absorption line outflows, which have been attributed to
starburst-driven winds, have been detected in many of the ULIRGs in
the aforementioned CO samples (Martin, 2005; Rupke et al.  2005a-c).

Apparent high velocity peaks appear prodigiously in \zsim 2
submillimeter Galaxies (SMGs; Greve et al. 2005; Tacconi et al.  2006;
see Blain et al., 2002 and Solomon \& Vanden Bout 2006 for recent
reviews).  If the high velocity peaks in SMGs owe to outflows, they
may be explained by the prodigious star formation rates in these
galaxies, in combination with a potential contribution from an
embedded AGN. For example, if the extreme luminosities owe solely to
starbursts, then massive star formation rates of order $\sim$1000
\msunyr are typical of these sources (e.g. Blain et al. 1999). The
relative role of AGN energy input in SMGs is a debated topic, as well
as whether they are currently undergoing, or are soon to undergo a
heavy black hole accretion phase (e.g. Alexander et al. 2005a,b; Borys
et al, 2005; Chakrabarti et al. 2007b). In either case, powerful winds
produced by the intense $\sim$10$^3$\msunyr SFR and/or the central
black hole could potentially drive high velocity peaks. That said, we
caution against a direct application of these model results toward
SMGs. For one thing, the virial properties of the progenitor galaxies
in these models have not been explicitly scaled for \zsim 2 (Robertson
et al. 2006a), and it is not clear how that will affect merger-induced
star formation rates and black hole growth during the active
period. Moreover, it is not completely clear that SMGs are the product
of binary mergers as this work has focused on, and the kinematic
profiles in these sources may be rather complex.  Models appropriate
for \zsim 2 mergers (e.g. Chakrabarti et al.  2007b; Robertson et
al. 2006a) will have to be studied in the context of galactic winds in
order to asses the potential relationship between these results and
SMGs.

\section{Conclusions and Summary}
\label{section:conclusions}

We have combined 3D non-LTE radiative transfer calculations with SPH
simulations of galaxy mergers in order to investigate the effects of
galactic-scale winds on the molecular line emission in starburst
galaxies and AGN. We find that galactic winds are a natural result of
merger-induced star formation and black hole growth. These winds can
entrain molecular gas of order $\sim$10$^8$-10$^9$ \msun which
imprints generic signatures in both the CO morphology as well as
unresolved emission line profiles. The specifics of the morphological
and emission line indicators of molecular outflows depend on physical
parameter choices within the galaxy merger models. In particular, the
energy source (i.e. BH accretion or star formation), as well as the
merger orientation can vary the strength, direction, and duration of
molecular outflows in emission contour maps, as well as the velocity
separation of high velocity peaks in multi-component emission
lines. Many of the results including the halflight radius as a
function of excitation, or the velocity offsets of high velocity peaks
in the emission lines vary enough between physical parameter choices
such that these models may be used to constrain physical origins for
observed CO morphologies and line profiles. In detail, we find the
following:

\begin{enumerate}

\item Molecular outflows entrained in AGN feedback and
starburst-driven winds can be directly imageable via CO emission line
mapping. These types of outflows have been recorded in at least some
local systems (e.g. Iono et al. 2007). These outflows will be
detectable at cosmological distances given the predicted spatial
resolution of ALMA.

\item Molecular outflows entrained in winds driven by AGN feedback are
typically longer lived than those entrained in starburst driven
winds. This owes to the relative strength of AGN feedback-driven winds
versus starburst driven winds.

\item Winds from AGN feedback in coplanar merger models are typically
more powerful than mergers which occur at more random
orientations. This results in outflows in these models being visible
for the majority of the $\sim$200 Myr 'active period' studied here.

\item The spatial extent of CO emission can be controlled by the
presence of galactic winds.  The emission is seen to be stratified
with transition in all models such that the CO (J=1-0) halflight
radius is larger than the radius from higher lying transitions
(e.g. CO J=7-6), though the degree of stratification depends on the
inclusion of winds.  The relatively compact nature of observed CO
emission in local mergers (R$_{1/2}$ typically confined to the central
kpc; Scoville \& Bryant, 1999) may be a consequence of AGN feedback or
starburst-driven winds.

\item In all models, high velocity peaks (peaks at velocities greater
than the circular velocity) can exist superposed on the post-merger
galaxy's broad CO emission line. In models without winds, these peaks
owe to random kinematics of molecular gas. In models with winds, these
peaks are seen to originate primarily from gas entrained in outflow,
at least during the period of peak black hole accretion/star
formation.

\item For the models studied here (Table~\ref{table:ICs}), high velocity
peaks driven by random kinematics do not typically appear at velocity
offsets (from systemic) greater than $\sim$1.7 times the circular
velocity of the post-merger galaxy. In contrast, peaks entrained in
AGN feedback-powered winds can be driven to velocities near 2.5 times
the circular velocity. The centroid velocities of the simulated lines
are typically (1$\sigma$) within $\sim$20 \kms of the true systemic
velocity, and can generally be used as a reliable substitute for the
systemic velocity in these models. Thus the above results hold true
when measuring the velocity offsets of high velocity peaks with
respect to line centroids as well.

\end{enumerate}

\acknowledgements We thank John Bieging, Sukanya Chakrabarti, Arjun
Dey, Chris Groppi, Fred Hamann, Daisuke Iono, Casey Papovich, Yancy
Shirley, and Linda Tacconi for helpful conversations. DN was funded
for this work by an NSF Graduate Research Fellowship.  The
calculations were performed at the Harvard-Smithsonian Center for
Parallel Astrophysical Computing. BR was supported in part by NASA
through the Spitzer Space Telescope Fellowship Program.  Support for
this work was also provided by NASA through grant number HST-AR-10308
from the Space Telescope Science Institute, which is operated by AURA,
Inc. under NASA contract NAS5-26555.

\end{document}